\documentclass{emulateapj}
\usepackage{graphicx}

\def\etal{et al.~}

\def\ba{\begin{eqnarray}}
\def\ea{\end{eqnarray}}

\begin{document}

\title{Planet formation in stellar binaries I: planetesimal dynamics in
massive protoplanetary disks}

\author{Roman R. Rafikov\altaffilmark{1} \& Kedron Silsbee\altaffilmark{1}}
\altaffiltext{1}{Department of Astrophysical Sciences,
Princeton University, Ivy Lane, Princeton, NJ 08540; 
rrr@astro.princeton.edu}


\begin{abstract}
About $20\%$ of exoplanets discovered by radial velocity surveys 
reside in stellar binaries. To clarify their origin
one has to understand the dynamics of planetesimals 
in protoplanetary disks within binaries. The standard description, 
accounting for only gas drag and gravity of the companion 
star has been challenged recently, as the gravity of the 
protoplanetary disk was shown to play a crucial role in 
planetesimal dynamics. An added complication
is the tendency of protoplanetary disks in binaries to become 
eccentric, giving rise to additional excitation of planetesimal 
eccentricity. Here, for the first time, we analytically explore 
secular dynamics of planetesimals in binaries such as $\alpha$ 
Cen and $\gamma$ Cep under the combined action of (1) 
gravity of the eccentric protoplanetary disk, (2) perturbations 
due to the (coplanar) eccentric companion, and (3) gas drag. 
We derive explicit solutions for the behavior of planetesimal 
eccentricity ${\bf e}_p$ in non-precessing disks (and in precessing 
disks in certain limits). We obtain the analytical form of the 
distribution of relative velocities of planetesimals, which is 
a key input for understanding their collisional evolution. 
Disk gravity strongly influences relative velocities and tends to 
push sizes of planetesimals colliding with comparable objects at 
the highest speed to small values, $\sim 1$ km. We also find 
that planetesimals in eccentric protoplanetary disks apsidally 
aligned with the binary orbit collide at lower relative 
velocities than in mis-aligned disks. Our results 
highlight a decisive role that disk gravity plays in 
planetesimal dynamics in binaries. 
\end{abstract}


\keywords{planets and satellites: formation --- 
protoplanetary disks --- planetary systems --- 
binaries: close}


\section{Introduction.}  
\label{sect:intro}


Results of radial velocity surveys demonstrate that 
$\sim 20\%$ of exoplanets reside in stellar binaries (Desidera 
\& Barbieri 2007). While most of these binaries have
wide separation between stellar components (hundreds of AU), 
some of them are systems with relatively short binary periods 
of $\sim 100$ yr. 
One of the best examples of such a binary is $\gamma$ Cephei 
(Hatzes \etal 2003), which consists of two stars of mass 
$M_p=1.6M_\odot$ and $M_s=0.41M_\odot$ with semi-major 
axis of $a_b=19$ AU (orbital period $P_b=58$ yr) and 
eccentricity $e_b=0.41$. The 
planet with the projected mass $M_{\rm pl}\sin i=1.6M_J$ 
is in orbit around the primary with semi-major axis 
$a_{\rm pl}\approx 2$AU. Several more planetary systems 
within tight ($a_b\approx 20$ AU) binaries are 
known at present (Chauvin \etal 2011), including the 
terrestrial planet around our stellar neighbor 
$\alpha$ Cen (Dumusque \etal 2012; cf. Hatzes 2013). 

For a long time theorists struggled to explain the origin of 
planets in such systems of S-type in classification of 
Dvorak (1982). The issue lies in the strong dynamical 
excitation that any object in a binary is subject to. 
Gravitational perturbations due to the eccentric companion 
are expected to adversely affect planet formation already 
at the stage of {\it planetesimal growth}. As first shown by 
Heppenheimer (1978) companion perturbations drive planetesimal 
eccentricities to high values, easily approaching $0.1$ at 2 
AU from the primary. Planetesimals would then be colliding 
at relative speeds of a couple km s$^{-1}$, which is 
much higher than the escape speed
from the surface of even a $100$ km object (about 100 m s$^{-1}$).
As a result, collisions should lead to planetesimal 
{\it destruction} rather than growth.

A number of possibilities have been explored to at least 
alleviate this problem. In particular, Marzari \& Scholl (2000),
studied the dissipative effects of {\it gas drag} as the means 
of damping relative velocities of planetesimals. These authors 
have shown that for a circular disk in secular approximation 
gas drag induces an 
{\it alignment} of planetesimal orbits such that the periapses of small 
objects strongly affected by gas drag tend to
cluster around $3\pi/2$ with respect to the binary apsidal line. 
This was originally thought (Marzari \& Scholl 2000; Th\'ebault 
\etal 2004) to assist planetesimal agglomeration 
since the relative velocities of colliding bodies are reduced 
by such orbital phasing. However, it was subsequently recognized 
(Th\'ebault \etal 2008) that the reduction of the relative velocity 
caused by apsidal alignment is effective only for 
planetesimals of similar sizes. Objects of different 
sizes still collide at high speeds, which complicates 
their growth.

These studies have generally arrived to the same conclusion --- 
difficulty of
planetesimal accretion --- despite the different ways in which the
gas disk and its interaction with planetesimals was treated. 
While the early calculations (Th\'ebault \etal 2004, 2006, 2008, 
2009; Th\'ebault 2011) typically assumed disk properties 
to be described by some (semi-)analytic axisymmetric models, 
recently several studies followed properties and evolution of gas 
disks in binaries using direct hydrodynamical simulations 
(Paardekooper \etal 2008; Kley \etal 2008; Marzari \etal 2009; 
Reg\'aly \etal 2011; M\"uller \& Kley 2012; Marzari \etal 2012; 
Picogna \& Marzari 2013). One of the most important aspects of 
the disk physics that the latter allow capturing is the 
{\it development of non-axisymmetry} in the surface density 
distribution of the gaseous disk. It emerges under the 
gravitational perturbation of the binary companion, predominantly in 
the form of non-zero eccentricity of the fluid trajectories 
(Marzari \etal 2012). Another phenomenon is the {\it disk 
precession}, with sometimes develops in simulations with 
subsequent effect on planetesimal dynamics.

An entirely different way of lowering planetesimal eccentricities 
in binaries has been pursued by Rafikov (2013a, 
hereafter R13), who demonstrated that planetesimal eccentricities 
can be considerably lower than previously thought by properly 
accounting for the gravity of a massive {\it axisymmetric} 
gaseous disk in which planetesimals form.  The non-Keplerian 
potential of the disk drives rapid precession of planetesimal 
orbits, suppressing driving of their eccentricity by 
the companion. 

Note that massive protoplanetary disks must have
been quite natural in $\gamma$ Cep-like systems since all of
the known systems (with the exception of $\alpha$ Cen) harbor 
Jupiter-like planets with $M_{\rm pl}\sin i=(1.6-4)M_J$ 
(Chauvin \etal 2011). It is natural to expect the parent disk 
mass to exceed the planet mass by at least a factor of 
several (this number is very uncertain but is believed to be 
$\sim 10$ for the Minimum Mass Solar Nebula) making an 
assumption of a $(0.01-0.1)M_\odot$ disk not unreasonable.
This is even despite the fact that sub-mm surveys find 
very low fluxes of thermal dust emission in young binaries 
with semi-major axes of several tens of AU (Harris \etal 2012).

As discussed above, the assumption of a purely 
axisymmetric disk may be too simplistic since simulations 
indicate that protoplanetary disks in binaries often 
develop significant eccentricities. To that effect Silsbee \& 
Rafikov (2013, hereafter SR13) presented the first investigation 
of secular excitation of planetesimal eccentricities by 
simultaneous action of the gravitational perturbations due to 
both the {\it eccentric} gaseous disk and the 
companion star. They showed that the non-axisymmetric 
gravitational field of such a disk excites planetesimal 
eccentricity (in addition to the excitation produced by the 
companion) and usually does not allow it to drop
below the disk eccentricity, which may be rather high as 
suggested by some simulations (Okazaki \etal 2002; Paardekooper 
\etal 2008; Kley \& Nelson 2008). This would again suppress 
planetesimal growth. On the other hand, SR13 outlined several 
ways in which this issue can be alleviated, for example if 
the gaseous disk is precessing rapidly or if its own self-gravity is capable
of reducing disk eccentricity to low levels. At the same time, 
SR13 did not include gas drag in their calculations, 
eliminating the possibility of planetesimal apsidal alignment 
and their eccentricity damping by drag.
 
Our current work builds upon the results of previous studies by 
exploring planetesimal dynamics in disks coplanar with the binary 
under the {\it combined effects} of (1) gravitational perturbations due to
the eccentric gaseous disk in which planetesimals are embedded,  
(2) gravity of eccentric companion, and (3) gas drag. While we do not 
model disks in binaries using hydrodynamical simulations, we 
still capture their main features, namely their non-axisymmetry 
and the possibility of precession. Our results are then used in 
a companion paper (Rafikov \& Silsbee 2014, in preparation; 
hereafter Paper II) to explore the details of planet formation 
in binaries.

We thereby extend the existing semi-analytical studies in 
which the disk is treated as axisymmetric with only gas drag 
(and not disk gravity) accounted for (Th\'ebault \etal 2004, 2006, 
2008, 2009). We also go beyond the works of R13 and SR13 in 
which gas drag was neglected and only the gravitational effects 
of the gaseous disk and binary companion were considered. In 
addition, we extend the study of Beaug\'e \etal (2010) 
devoted to exploring planetesimal dynamics in eccentric, 
precessing disks by accounting for the gravitational 
potential of such a disk. 

This paper is structured as follows. We discuss our general 
setup in \S \ref{sect:setup} and then derive equations for 
the evolution of orbital elements of planetesimals in eccentric 
disks in \S \ref{sect:basic_eqs}. Prescription for the gas 
drag-induced eccentricity evolution is described in 
\ref{sect:drag}. Solutions of equations
of planetesimal dynamics in non-precessing and precessing disks 
are presented in \S \ref{sect:ecc_evol} and \ref{sect:precess} 
(as well as Appendices \ref{sect:lin_t_sol} \&
\ref{sect:quad_t_sol}) correspondingly. The diversity of 
planetesimal dynamical behaviors is discussed in  \S 
\ref{sect:pl_dyn}. We derive the relative 
velocity distribution of objects of different sizes in \S 
\ref{sect:rel_vel}. We provide an extensive 
discussion of our dynamical results and their applications 
in \S \ref{sect:disc}. We compare different approximations for
treating planetesimal dynamics in \S \ref{sect:dyn_approx},
briefly discuss limitations of this work 
in \S \ref{sect:limits} and summarize our main conclusions in \S   
\ref{sect:summ}. Finally, some of our analytical derivations use the 
local approximation for treating elliptical motion, which is 
reviewed in Appendix \ref{sect:local}.


\section{Problem setup.}  
\label{sect:setup}


Our general setup is similar to that explored in SR13. We
consider an elliptical disk around a primary star in a binary 
with semi-major axis $a_b$, eccentricity $e_b$, and component 
masses $M_p$ (primary) and $M_s$ (secondary). We define 
$\mu\equiv M_s/(M_p+M_s)$ and $\nu\equiv M_s/M_p$. Binary, 
disk, and planetesimal orbits within it are assumed to be 
{\it coplanar}. This distinguishes our work from many other studies 
focused on the effects of Lidov-Kozai oscillations (Lidov 1962; 
Kozai 1962) on planetesimal dynamics in systems with inclined 
companions (Marzari \etal 2009; Batygin \etal 2011; Zhao \etal 2012).

Non-axisymmetric disk structure is described via the non-zero 
disk eccentricity, which is a viable approximation given that 
simulations tend to show the prevalence of $m=1$ azimuthal 
harmonic of the disk shape distortion (Marzari \etal 2012). 
Fluid elements in a disk follow 
elliptical trajectories with eccentricity $e_g(a_d)$ which is 
a function of the semi-major axis $a_d$ of a particular 
ellipse. All of them have the primary star of the binary as a
focus. For simplicity we assume all fluid elliptical trajectories 
to have {\it aligned} apsidal lines, uniquely determining disk 
orientation via a single parameter $\varpi_d$ --- the angle 
between the disk and binary apsidal lines. The latter is 
assumed to be fixed in space as the precession of 
the binary under the gravity of the disk is slower than all
other processes. The assumption of apsidal alignment 
does not affect the qualitative features of planetesimal 
dynamics and can be easily relaxed using the results of Statler 
(2001). 

Because gas moves on ellipses its surface density generally 
varies along the trajectory (Statler 2001; Ogilvie 2001).
To obtain gas surface density $\Sigma(r_d,\phi_d)$ at a point 
in a disk with polar coordinates $(r_d,\phi_d)$ we specify 
gas surface density at periastron of each elliptical trajectory
$\Sigma_p(a_d)$ as a function of semi-major axis of the 
corresponding ellipse $a_d$. SR13 show how this and the knowledge 
of $e_g(a_d)$ can be used to derive $\Sigma(r_d,\phi_d)$ everywhere 
in the disk. In this work, following SR13, we assume simple power
law dependence for both $e_g$ and $\Sigma_p$:
\ba
\Sigma_p(a_d)=\Sigma_0\left(\frac{a_{\rm out}}{a_d}\right)^p,~~~
e_g(a_d)=e_0\left(\frac{a_{\rm out}}{a_d}\right)^q.
\label{eq:Sigma0}
\ea
where $a_{\rm out}$ is the semi-major axis of the outermost 
elliptical trajectory of the disk, and $\Sigma_0$ and
$e_0$ are the values of $\Sigma_p$ and $e_g$ at $a_{\rm out}$.
Gravity of the companion 
truncates the disk at this outer radius $a_{\rm out}$, which
for eccentric binaries with $e_b=0.4$ is about $(0.2-0.3)a_b$ 
(Artymowicz \& Lubow 1994; Reg\'aly \etal 2011). Unless stated 
otherwise we will be using $a_{\rm out}=5$ AU in this work.

In all calculations of this paper we 
will be using a disk model with $p=1$ and $q=-1$. Some
 motivation for singling out these particular 
values of $p$ and $q$ for circumstellar disks in binaries 
has been provided in R13 and SR13. 

The total disk mass $M_d\approx 2\pi 
\int^{a_{\rm out}}_{a_{\rm in}}\Sigma_p(a_d)a_d da_d$ 
enclosed within $a_{\rm out}$ can be used to express 
$\Sigma_p$ as
\ba
\Sigma_p(a_d) & = & \frac{2-p}{2\pi}\frac{M_d}{a_{\rm out}^2}
\left(\frac{a_{\rm out}}{a_d}\right)^p
\label{eq:sig_0}\\
& \approx & 3\times 10^3~
\mbox{g cm}^{-2}M_{d,-2}a_{\rm out,5}^{-1}a_{d,1}^{-1},
\nonumber
\ea
(numerical estimate is for $p=1$) where 
$M_{d,-2}\equiv M_d/(10^{-2}M_\odot)$, $a_{\rm out,5}\equiv a_{\rm out}/(5$ 
AU) and $a_{d,1}\equiv a_d/$AU. In equation (\ref{eq:sig_0}) 
we neglected disk ellipticity and assumed $p<2$, so that most 
of the disk mass is concentrated in its outer part.


\section{Basic equations.}  
\label{sect:basic_eqs}

We are interested in the dynamics of planetesimals orbiting the 
primary within the disk and coplanar with it. We characterize 
their orbits by semi-major axis $a_p$, eccentricity $e_p$ and the 
apsidal angle (w.r.t. the binary apsidal line) $\varpi_p$. Orbital 
evolution of planetesimals is
treated in secular approximation, i.e. neglecting short-term 
gravitational perturbations (Murray \& Dermott 1999). We 
also assume $e_p\ll 1$ as well as $e_g\ll 1$ and introduce for 
convenience the planetesimal eccentricity vector ${\bf e}_p=(k_p,h_p)=
e_p(\cos\varpi_p,\sin\varpi_p)$. 

In this work we fully account for gravitational perturbations 
due to both the binary companion and the eccentric disk using 
the approach advanced in SR13. For the disk properties described 
by equation (\ref{eq:Sigma0}) SR13 calculated an analytic expression 
for the planetesimal disturbing function accounting for the gravity 
of both disk and secondary. They then derived a set of Lagrange
equations [see their equations (16)-(17)] describing the evolution 
of ${\bf e}_p$ under the influence of gravitational forces alone.

In addition, in this work we take 
into account the effects of gas drag on the secular evolution of 
planetesimal eccentricity. Drag-induced dissipation also results in 
non-conservation of energy and evolution of $a_p$. However, to 
zeroth order we can neglect this as the radial inspiral of 
planetesimals usually occurs on much longer timescale than their 
eccentricity evolution (Adachi \etal 1976). As a result, we 
can concentrate on 
the behavior of ${\bf e}_p$ at fixed $a_p$ and 
determine the relative velocities of planetesimals and their 
collisional outcomes.

Gas drag introduces additional terms in the eccentricity evolution 
equations of SR13, which we re-write in the following form:
\ba
&&\frac{dh_p}{dt}=Ak_p+B_b+B_d\cos\varpi_d(t)+\dot h_p^{\rm drag},
\label{eq:dhdt_gen}\\
&&\frac{dk_p}{dt}=-Ah_p-B_d\sin\varpi_d(t)+\dot k_p^{\rm drag}.
\label{eq:dkdt_gen}
\ea
Here $A=A_b+A_d$ is the planetesimal precession rate. It is 
contributed both by the gravity of the secondary ($A_b$) and 
the disk ($A_d$), with
\ba
A_b & = & \frac{\nu}{4}n_p \alpha_b^2 b_{3/2}^{(1)}(\alpha_b)\approx 
\frac{3}{4}n_p\nu \left(\frac{a_p}{a_b}\right)^3
\label{eq:A_b}\\
& \approx & 5.9\times 10^{-4}\mbox{yr}^{-1}\nu 
\frac{M_{p,1}^{1/2}}{a_{b,20}}a_{p,1}^{3/2},
\nonumber
\ea
where $n_p=(GM_p/a_p^3)^{1/2}$ is the planetesimal mean rate, 
$M_{p,1}\equiv M_{p}/M_\odot$, $a_{p,1}\equiv a_{p}/$AU, 
$a_{b,20}\equiv a_{b}/(20$ AU), 
$b_s^{(j)}(\alpha)$ is the standard Laplace coefficient 
(Murray \& Dermott 1999), $\alpha_b\equiv a_p/a_b$ and the 
approximation in (\ref{eq:A_b}) works for $\alpha_b\ll 1$. 
The disk contribution is 
\ba
A_d & = & 2\pi\frac{G \Sigma_p(a_p)}{a_p n_p}\psi_1=
(2-p)\psi_1 n_p\frac{M_d}{M_p}\left(\frac{a_p}{a_{\rm out}}
\right)^{2-p}
\label{eq:A_d}\\
& \approx & -6.3\times 10^{-3}\mbox{yr}^{-1}a_{p,1}^{-1/2}
\frac{M_{d,-2}}{M_{p,1}^{1/2}a_{\rm out,5}}
\nonumber
\ea
where the numerical 
estimate is for $p=1$ so that $\psi_1=-0.5$ (SR13).
Dimensionless coefficients of order unity $\psi_1$ and 
$\psi_2$ (see equation 
(\ref{eq:B_d})) have been calculated in SR13 and are functions of 
the disk model and the distance of planetesimal orbit from the disk edges.
One can see that for reasonable assumptions about the disk mass 
($M_d\sim 10^{-2}M_\odot$) the planetesimal precession rate at 1 AU is
dominated by the disk gravity. 

Eccentricity {\it excitation} by the binary ($B_b$) and the disk 
($B_d$) are described by 
\ba
B_b & = & - \frac{\nu}{4}n_p\alpha_b^2 b_{3/2}^{(2)}(\alpha_b)e_b\approx 
-\frac{15}{16}n_p\nu \left(\frac{a_p}{a_b}\right)^4e_b,
\label{eq:B_b}\\
B_d & = & \pi \frac{G \Sigma_p(a_p)}{a_p n_p}e_d(a_p)\psi_2
\label{eq:B_d}\\
& = &
\frac{2-p}{2}\psi_2 e_g(a_p)n_p\frac{M_d}{M_p}\left(\frac{a_p}
{a_{\rm out}}\right)^{2-p},
\ea
with the latter explicitly depending on the local value of the disk 
eccentricity $e_g(a_p)$.

Note that $\varpi_d$ in equations (\ref{eq:dhdt_gen}-\ref{eq:dkdt_gen}) 
is not necessarily constant --- it can be an explicit 
function of time, allowing one to treat the case of a precessing 
disk. 

Terms $\dot h_p^{\rm drag}$ and $\dot k_p^{\rm drag}$ absent 
in the original version of equations 
(\ref{eq:dhdt_gen})-(\ref{eq:dkdt_gen}) in SR13 represent the
effect of gas drag on the eccentricity evolution; they are 
derived in \S \ref{sect:drag}. The main goal of this work is to 
see how their introduction affects planetesimal dynamics.


\section{Drag force calculation.}  
\label{sect:drag}


Next we derive the expressions for the drag-induced eccentricity 
evolution terms $\dot h_p^{\rm drag}$ and $\dot k_p^{\rm drag}$
applicable to the case of an eccentric disk. 

Because of our assumption of small eccentricities for both 
gas and planetesimals, it is reasonable to employ the local (or 
guiding center) approximation. This approach is often used in 
studies of planetesimal and galactic dynamics (Binney \& 
Tremaine 2008) and forms a basis of the so-called Hill 
approximation (H\'enon \& Petit 1986; Hasegawa \& Nakazawa 1990). 
It considers planetesimal motion in a local Cartesian 
$x-y$ reference system aligned with radial and
azimuthal directions, respectively. Main features of 
this approximation are reviewed in Appendix 
\ref{sect:local}. In particular, equations (\ref{eq:kh_dot}) 
describe how $k_p$ and $h_p$ evolve under the effect of 
external force ${\bf F}$.

In our case ${\bf F}$ is the drag force arising because of 
the motion of planetesimals with respect to gas. Adachi 
\etal (1976) gives the following expression for quadratic 
drag force appropriate for rapidly moving objects with size 
larger than the mean free path of gas molecules:
\ba
{\bf F}=-\frac{C_D}{2}\pi d_p^2\rho_g v_r{\bf v}_r,
\label{eq:F_drag}
\ea
 where $C_D$ is a constant drag coefficient taken to be $0.5$ 
throughout this paper, $d_p$ is the 
particle size, and $\rho_g$ is the gas density. The relative
particle-gas velocity ${\bf v}_r$ is given by equations 
(\ref{eq:v_unpertxy}), (\ref{eq:v_unperty1}), and 
(\ref{eq:v_unper}) with relative particle-gas eccentricity 
components
\ba
h_r=h_p-h_g,~~~k_r=k_p-k_g,
\label{eq:rel_gas}
\ea
and ${\bf e}_g=(k_g,h_g)=e_g(\cos\varpi_d,\sin\varpi_g)$ 
being the local value of the gas eccentricity vector. Using 
these expressions we obtain the force components $F_x$
and $F_y$:
\ba
F_x & =& -\frac{3C_D}{8}m_p D ~v_r^a
\left(k_r\sin n_p t - h_r\cos n_p t\right),
\label{eq:F_x}\\
F_y & =& -\frac{3C_D}{16}m_p D ~v_r^a
\left(k_r\cos n_p t + h_r\sin n_p t\right),
\label{eq:F_y}
\ea
where $m_p$ is the planetesimal mass, and the relative velocity 
$v_r^a$ is given by equation (\ref{eq:v_unper}). The 
prefactor $D$ is given by 
\ba
D=n_p\frac{\Sigma_g}{\rho_p d_p}\frac{r}{h},
\label{eq:D}
\ea
with $\rho_p$ being the particle bulk density and $h=c_s/n_p$ 
being the disk scale height ($c_s=\left(kT_g/\mu\right)^{1/2}$).

Now we plug the expressions for $F_x$, $F_y$ into the first two 
equations (\ref{eq:kh_dot}) and then average them in time $t$ over 
planetesimal orbital period (this is the secular, i.e. time-averaged 
approximation). One can easily see that to get the result 
to lowest-order in $e_r$ we do not need to keep terms $O(e_r,e_d,e_p)$
in the expression for $\rho_g$. As a result we find 
\ba
&& \dot k_p^{\rm drag}=-\frac{3C_D}{4\pi}\mbox{E}\left(\frac{\sqrt{3}}{2}
\right)D ~k_r e_r,
\label{eq:kdrag}\\
&& \dot h_p^{\rm drag}=-\frac{3C_D}{4\pi}\mbox{E}\left(\frac{\sqrt{3}}{2}
\right)D ~h_r e_r,
\label{eq:hdrag}
\ea
where $E\left(\sqrt{3}/2\right)\approx 1.211$ is a complete elliptic 
integral, and $e_r^2=k_r^2+h_r^2$.

We can rewrite equations (\ref{eq:kdrag})-(\ref{eq:hdrag}) 
in the following form:
\ba
\dot k_p^{\rm drag}=-\frac{k_p-k_g}{\tau_d},~~~~~~
\dot h_p^{\rm drag}=-\frac{h_p-h_g}{\tau_d},
\label{eq:hdrag1}
\ea
where the {\it eccentricity damping time} 
\ba
\tau_d & = & \frac{4\pi}{3C_D\mbox{E}\left(\sqrt{3}/2\right)}D^{-1}e_r^{-1}
\label{eq:tau_d}\\
& \approx & 600~\mbox{yr}~C_D^{-1}
\frac{a_{\rm out,5}a_{p,1}}{M_{p,1}^{1/2}M_{d,-2}}
\frac{h/r}{0.1}\frac{10^{-2}}{e_r}d_{p,1}.
\nonumber
\ea
Here $d_{p,1}\equiv d_p/(1$ km) and numerical estimate is for $p=1$
and $\rho_p=3$ g cm$^{-3}$; in the case of 
quadratic drag law (\ref{eq:F_drag}) $\tau_d$
depends on $k_p$ and $h_p$ through $e_r$, see equation (\ref{eq:rel_gas}).

\begin{figure}
\epsscale{1.3}
\plotone{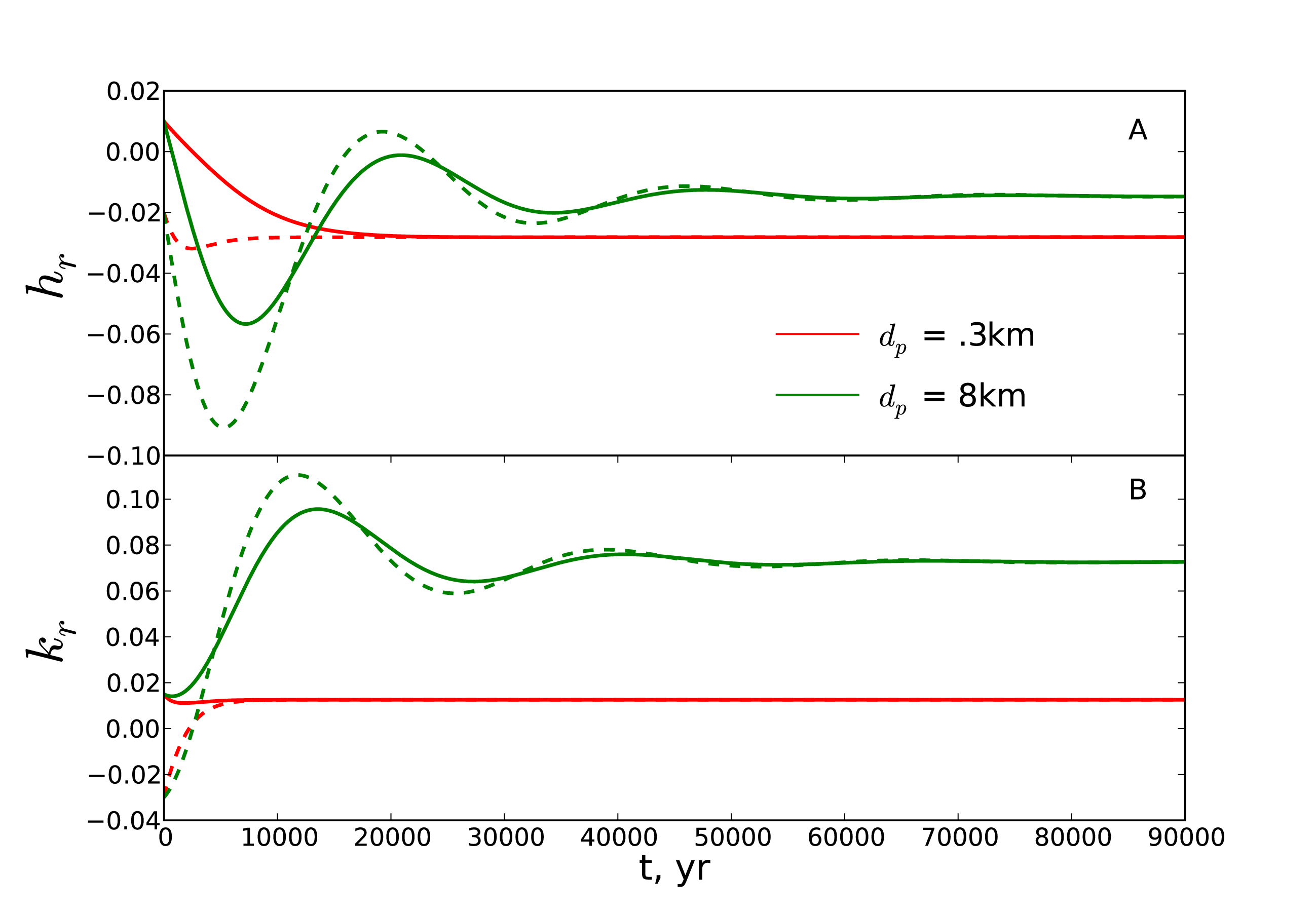}
\caption{
Time evolution of the components of the relative eccentricity vector 
${\bf e}_r=(k_r,h_r)={\bf e}_p-{\bf e}_g$ for planetesimals of two 
different sizes --- $d_p=0.3$ km (red) and 8 km (green) starting 
with two different sets of initial conditions --- ${\bf e}_r=(0.015,0.01)$
(solid) and ${\bf e}_r=(-0.03,-0.02)$ (dashed). Calculations are 
carried out for the parameters of $\gamma$ Cep system at 2 AU in a 
$0.001M_\odot$ disk aligned with the binary; disk eccentricity at 
its outer edge $a_{\rm out}=5$ AU is $e_0=0.05$, and $p=1$, $q=-1$, 
see equations (\ref{eq:Sigma0}). After the short initial transient 
$k_r$ and $h_r$ 
converge to the forced values given by equations 
(\ref{eq:k_p})-(\ref{eq:h_p}).
\label{fig:time_ev}}
\end{figure}


\section{Eccentricity evolution.}  
\label{sect:ecc_evol}


Results of \S \ref{sect:drag} allow us to understand the 
behavior of ${\bf e}_p$. For simplicity, we start by considering 
the case of a non-precessing
disk, i.e. $\varpi_d=$const. Even in this case equations 
(\ref{eq:dhdt_gen})-(\ref{eq:dkdt_gen}) with the 
quadratic drag terms (\ref{eq:kdrag})-(\ref{eq:hdrag}) 
cannot be solved analytically in general because of the 
$\tau_d$ dependence on $e_r$. 

\begin{figure}
\vspace{-1cm}
\epsscale{1.2}
\plotone{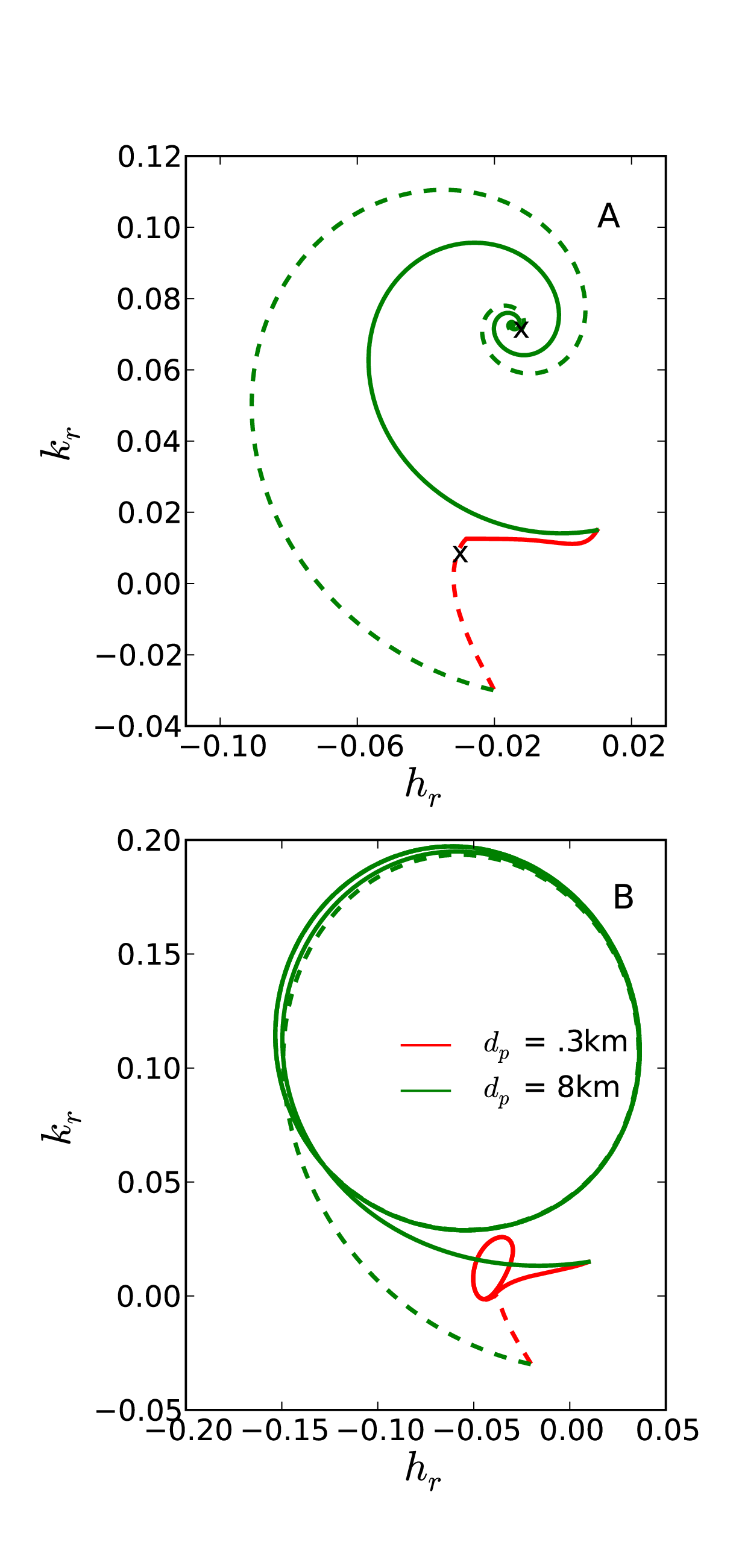}
\caption{
 (a) Planetesimal eccentricity evolution trajectories in $k_r-h_r$ 
space in a non-precessing, aligned ($\varpi_d=0$) disk for the 
four cases shown in Figure \ref{fig:time_ev}. One can see the 
convergence of trajectories starting at different ${\bf e}_p$ 
to fixed point solutions (indicated by crosses), which 
depend on planetesimal radius $d_p$. (b) Same for a 
disk precessing at the rate of $\dot\varpi_d=A$. Evolution 
trajectories converge to a limit cycle behavior in the 
precessing disk. See \S \ref{sect:precess} for more 
details. Color and line type scheme is the same for both 
panels.
\label{fig:ev_traj}}
\end{figure}

However, it can be easily shown that solutions of these equations 
inevitably converge to a steady-state form --- the free 
eccentricity, which depends on initial conditions (R13, SR13), damps 
out and ${\bf e}_p$ converges to the forced eccentricity 
vector (Beaug\'e et al 2010). This is illustrated in Figure 
\ref{fig:time_ev} where we solve evolution equations numerically. 
It is clear that starting with 
arbitrary initial conditions and after initial (sometimes 
oscillatory) evolution $k_p$ and $h_p$ do converge to the 
same steady state values (depending only on the disk parameters 
and planetesimal size), which are given by equations 
(\ref{eq:k_p})-(\ref{eq:h_p}) derived below.
This point is additionally illustrated in Figure \ref{fig:ev_traj}a
where we plot the trajectory of ${\bf e}_p$ as it evolves in
$h_r$-$k_r$ coordinates. There one can clearly see ${\bf e}_p$
converging to a {\it fixed point} solution, in oscillatory 
fashion for large planetesimals, and exponentially for 
small objects, which rapidly couple to the gas disk.

Damping of the memory of initial conditions can also be
demonstrated by solving equations (\ref{eq:dhdt_gen})-(\ref{eq:dkdt_gen})
analytically in a simplified but qualitatively similar case 
of a {\it linear} drag law, when $\tau_d$ is independent of 
$h_p$ and $k_p$. Such solution is presented in Appendix 
\ref{sect:lin_t_sol} for the general case of a precessing gaseous 
disk. Non-precessing disk solution is obtained by setting 
$\dot \varpi_d=0$. It clearly demonstrates the convergence 
of ${\bf e}_p$ to a time-independent, forced value.

SR13 have demonstrated that {\it in the absence gas drag}, under 
the action of only the gravity of the disk and the companion star, 
the steady state (forced) eccentricity is given by 
\ba
{\bf e}_p^{\rm n/drag} & = & 
\left\{
\begin{array}{l}
k_p^{\rm n/drag}\\
h_p^{\rm n/drag}
\end{array}
\right\}=
{\bf e}_{b}+{\bf e}_{d},
\label{eq:e_n_drag_decompose}\\
{\bf e}_{b} & = & 
\left\{
\begin{array}{l}
k_b\\
h_b
\end{array}
\right\}=
-\frac{B_b}{A}
\left\{
\begin{array}{l}
1 \\
0
\end{array}
\right\},
\label{eq:e_forced_b}\\
{\bf e}_{d} & = & 
\left\{
\begin{array}{l}
k_d\\
h_d
\end{array}
\right\}=
-\frac{B_d}{A}
\left\{
\begin{array}{l}
\cos\varpi_d \\
\sin\varpi_d
\end{array}
\right\},
\label{eq:e_forced_d}
\ea
where ${\bf e}_{b}$ and ${\bf e}_{d}$ are forced 
eccentricity vectors due to the secondary and disk gravity,
respectively. Note that the accuracy of analytical expression 
(\ref{eq:e_forced_b}) for the binary contribution is known 
to worsen (beyond the $\sim 10\%$ level) when 
$a_p/a_b\gtrsim 0.1$ (Th\'ebault \etal 2006; Barnes \& 
Greenberg 2006). 
More refined calculations of ${\bf e}_{b}$ are possible 
(Veras \& Armitage 2007; Giuppone \etal 2011) but for the 
purposes of this work it is sufficient to use equation 
(\ref{eq:e_forced_b}).

With the gas drag included the behavior of ${\bf e}_p$ changes.
To determine the steady-state values of $k_p$ and $h_p$ and 
analyze their properties we use the prescription 
(\ref{eq:hdrag1}), set to zero time derivatives in the 
left-hand-sides of equations (\ref{eq:dhdt_gen})-(\ref{eq:dkdt_gen}) 
and solve the resulting algebraic system with respect to $h_p$ and 
$k_p$. We find as a result 
\ba
k_p & = & k_b+k_d + \frac{
\left(k_g-k_b-k_d\right) - \left(h_g-h_d\right)\left(A\tau_d\right)
}{1+\left(A\tau_d\right)^2},
\label{eq:k_p}\\
h_p & = & h_d+ \frac{
\left(h_g - h_d\right) + \left(k_g-k_b-k_d\right)\left(A\tau_d\right)}
{1+\left(A\tau_d\right)^2},
\label{eq:h_p}
\ea
where $k_b$, $k_d$, $h_b$, $h_d$ are defined in equations 
(\ref{eq:e_forced_b})-(\ref{eq:e_forced_d}). These asymptotic 
results are valid even {\it if $\tau_d$ is a function of $e_r$} 
--- in that case they simply represent two implicit relations 
for $h_p$ and $k_p$. 

Solutions (\ref{eq:k_p})-(\ref{eq:h_p}) can be re-written 
 in vectorial form as
\ba
{\bf e}_p & = &  \left\{
\begin{array}{l}
k_p\\
h_p
\end{array}
\right\}={\bf e}_{f,b}+{\bf e}_{f,d},
\label{eq:vect_sum}
\\
{\bf e}_{f,b} & = &
k_b\frac{(A\tau_d)}{1+(A\tau_d)^2}
\left\{
\begin{array}{l}
(A\tau_d)\\
-1
\end{array}
\right\},
\label{eq:e_f_b}\\
{\bf e}_{f,d} & = & 
\left[\frac{e_g^2+\tau_d^2 B_d^2}{1+(A\tau_d)^2}\right]^{1/2}
\left\{
\begin{array}{l}
\cos\left(\varpi_d+\phi\right)\\
\sin\left(\varpi_d+\phi\right)
\end{array}
\right\},
\label{eq:e_f_d}
\ea
where the phase shift $\phi$ is given by
\ba
\cos\phi=\frac{e_g- A B_d\tau_d^2}
{\left(e_g^2+\tau_d^2 B_d^2\right)^{1/2}
\left[1+(A\tau_d)^2
\right]^{1/2}}.
\label{eq:phi_nonprec}
\ea
In the limit of vanishing drag, $A\tau_d\to \infty$, one finds 
$\phi\to \pi$ and solution (\ref{eq:vect_sum})-(\ref{eq:e_f_d})
reduces to the non-drag result with no free eccentricity
(\ref{eq:e_n_drag_decompose})-(\ref{eq:e_forced_d}), see SR13. 

In the limit of strong drag ($A\tau_d\to 0$) in a circular disk 
(i.e. $e_g=0$) and no disk gravity (i.e. $A_d=B_d=0$) one 
finds $h_p/k_p\to -\infty$. This means that in this case 
planetesimal apsidal lines cluster around $\varpi_p=3\pi/2$, 
in agreement with Marzari \& Scholl (2000). Also, 
$|{\bf e}_p|\to B_b\tau_d$ directly depends on 
planetesimal size, which implies that in this limit 
planetesimals of different sizes collide with non-zero
speeds even despite their apsidal alignment 
(Th\'ebault \etal 2008).

Expressions (\ref{eq:vect_sum})-(\ref{eq:e_f_d}) clearly show that 
${\bf e}_p$ can be split into two distinct components: a 
contribution ${\bf e}_{f,b}$ due to the gravity of the binary 
and a contribution ${\bf e}_{f,b}$ related to both the 
gravitational and gas drag effects of the disk. It is also 
clear that after reaching steady state planetesimal orbits 
are in general aligned with neither the disk ($\varpi_p\neq 
\varpi_d$) nor the binary ($\varpi_p\neq 0$).


\subsection{Relative particle-gas eccentricity.}  
\label{sect:rel_ec}

In the case of quadratic drag (\ref{eq:F_drag}) we can further 
analyze eccentricity behavior. Using equations 
(\ref{eq:k_p})-(\ref{eq:h_p}) we express relative particle-gas 
eccentricity as
\ba
e_r=\left|{\bf e}_p - {\bf e}_g\right|=
e_c\frac{\left(A\tau_d\right)}{\sqrt{1+\left(A\tau_d\right)^2}},
\label{eq:e_r}
\ea
where we introduced a characteristic eccentricity 
$e_c=\left|{\bf e}_p^{\rm n/drag} - {\bf e}_g\right|$ given by 
\ba
e_c & \equiv & \left[\left(h_g-h_d\right)^2+
\left(k_g-k_b-k_d\right)^2\right]^{1/2}
\label{eq:e_c}\\
& = &
\frac{\left[(Ae_g+B_d)^2+B_b^2+2\cos\varpi_dB_b(Ae_g+B_d)\right]^{1/2}}
{|A|}.
\nonumber
\ea

Plugging this expression for $e_r$ into equation (\ref{eq:tau_d})
one obtains the following bi-quadratic equation for 
$\left(A\tau_d\right)$:
\ba
\left(A\tau_d\right)^4=\left(\frac{d_p}{d_c}\right)^2
\left[\left(A\tau_d\right)^2+1\right],
\label{eq:quad}
\ea
where we have introduced a characteristic planetesimal size $d_c$
defined as
\ba
d_c \equiv \frac{3C_D\mbox{E}
\left(\sqrt{3}/2\right)}{4\pi}\frac{n_p}{|A|}\frac{\Sigma_g}{\rho_p}
\frac{r}{h}e_c.
\label{eq:d_1}
\ea
All our subsequent results can be formulated 
completely in terms of $e_c$ and $d_p/d_c$, underscoring
the significance of these variables. Detailed discussion of the
characteristic values and general behavior of $e_c$ and $d_c$ is 
provided in \S \ref{sect:e_c} and \ref{sect:d_c}.

Solving equation (\ref{eq:quad}) one finds
\ba
\left|A\tau_d\right|=\frac{d_p}{d_c}\left[\frac{1}{2}+
\sqrt{\frac{1}{4}+\left(\frac{d_c}{d_p}\right)^2}\right]^{1/2},
\label{eq:Atd}
\ea
i.e. that $\left|A\tau_d\right|$ is a function of $d_r/d_c$ only.

Plugging (\ref{eq:Atd}) into (\ref{eq:e_r}) one also finds the
general expression for the relative particle-gas eccentricity
\ba
e_r=e_c\frac{d_p}{d_c}
\left[\sqrt{\frac{1}{4}+\left(\frac{d_c}{d_p}\right)^2}
-\frac{1}{2}\right]^{1/2}
\label{eq:e_r_d}
\ea
valid for arbitrary $d_p/d_c$.

\begin{figure}
\epsscale{1.2}
\plotone{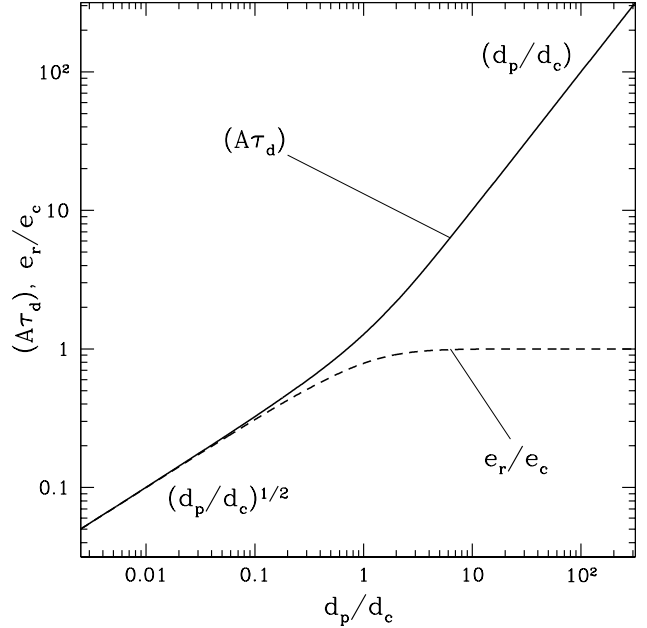}
\caption{
Dependence of $e_r/e_c$ and $\left|A\tau_d\right|$ on planetesimal 
size $d_p/d_c$, given by equations (\ref{eq:e_r}) and (\ref{eq:Atd}) 
respectively. Asymptotic scalings (\ref{eq:Atd_s}) and 
(\ref{eq:Atd_w}) are also indicated. For $d_p\sim d_c$ one finds
$\left|A\tau_d\right|\sim 1$ and $e_r\sim e_c$.
\label{fig:sols}}
\end{figure}

We illustrate the behaviors of $\left|A\tau_d\right|$ and $e_r$ 
given by equations (\ref{eq:Atd}) and (\ref{eq:e_r_d}) in 
Figure \ref{fig:sols}.  It reveals the meaning of the 
characteristic size $d_c$: objects with $d_p\sim d_c$ have 
$\left|A\tau_d\right|\sim 1$, i.e. their stopping time due to gas 
drag is comparable to their orbital precession period, and 
their relative eccentricity with respect to gas is $e_r\sim e_c$. 

It is instructive to further explore general solutions 
(\ref{eq:Atd}), (\ref{eq:e_r_d}) valid for arbitrary $d_p/d_c$ 
in the two limits covered next. 

\subsection{Small objects, $d_p\lesssim d_c$ --- strong drag 
($|A\tau_d|\lesssim 1$).}  
\label{sect:strong}

In the limit of {\it strong} gas drag we expect damping time 
$\tau_d$ to be very short and $|A\tau_d|\ll 1$, so that gas-particle 
velocity differential is rapidly reduced to zero. According to 
equation (\ref{eq:Atd}), this regime is valid for small objects 
with $d_p\lesssim d_c$, when
\ba
\left|A\tau_d\right|\approx \left(d_p/d_c\right)^{1/2}\lesssim 1.
\label{eq:Atd_s}
\ea
From equation (\ref{eq:e_r}) the relative particle-gas 
eccentricity is
\ba
e_r \approx \left|A\tau_d\right| e_c\approx 
e_c\left(d_p/d_c\right)^{1/2}
\label{eq:e_r_s}
\ea
to leading order in $\left(A\tau_d\right)$.

Equations (\ref{eq:k_p})-(\ref{eq:h_p}) become
\ba
k_p\to k_g & + & \big[\left(h_d-h_g\right)\left(A\tau_d\right)
\nonumber\\
& - & \left(k_g-k_b-k_d\right)\left(A\tau_d\right)^2 \big],
\label{eq:k_p_s}\\
h_p\to h_g & + & \big[\left(k_g-k_b-k_d\right)\left(A\tau_d\right)
\nonumber\\
& + & \left(h_d-h_g\right)\left(A\tau_d\right)^2 \big].
\label{eq:h_p_s}
\ea
Here brackets encompass the leading order subdominant terms, 
compared to the zeroth order terms outside brackets. 

It is clear from these asymptotic expressions that in the case 
of strong drag, the eccentricity vector of planetesimals tends to 
the eccentricity vector of the gas, 
${\bf e}_p\to {\bf e}_g$.  It is only 
weakly sensitive to gravitational perturbations due to either
the companion or the disk. Thus, to leading order the value of 
eccentricity vector is {\it independent of particle size} 
(which enters only through $\tau_d$).


\subsection{Big objects, $d_p\gtrsim d_c$ ---  weak drag 
($|A\tau_d|\gtrsim 1$).}  
\label{sect:weak}

In the opposite limit of {\it weak} drag or long damping time 
$|A\tau_d|\gg 1$ valid for large objects with $d_p\gtrsim d_c$
equation (\ref{eq:Atd}) yields
\ba
\left(A\tau_d\right)\approx d_p/d_c\gtrsim 1,
\label{eq:Atd_w}
\ea
while the relative particle-gas eccentricity is
\ba
e_r \approx e_c
\label{eq:e_r_w}
\ea
see equation (\ref{eq:e_r}). Thus, in the weak drag regime $e_r$
saturates at the value independent of the size of the object.

Equations (\ref{eq:k_p})-(\ref{eq:h_p}) reduce in this limit to 
\ba
k_p\to k_b+k_d & + & \big[\left(h_d-h_g\right)\left(A\tau_d\right)^{-1}
\nonumber\\
& + & \left(k_g-k_b-k_d\right)\left(A\tau_d\right)^{-2} \big],
\label{eq:k_p_w}\\
h_p\to h_g & + & \big[\left(k_g-k_b-k_d\right)\left(A\tau_d\right)^{-1}
\nonumber\\
& - & \left(h_d-h_g\right)\left(A\tau_d\right)^{-2} \big].
\label{eq:h_p_w}
\ea
Again, terms in brackets are subdominant compared to
the leading terms (outside brackets). 

This solution shows that in the limit of weak drag 
${\bf e}_p\to {\bf e}_p^{\rm n/drag}$, i.e. the behavior of the 
particle eccentricity vector is determined predominantly 
by the gravitational effects of the secondary and the disk. 
Thus, ${\bf e}_p$ is again almost independent of the particle 
size.


\section{Precessing disks.}  
\label{sect:precess}


So far we have assumed the orientation of the disk to be fixed 
in the binary frame. However, some simulations find
disks in binaries to precess (e.g. Marzari et al 2009; 
M\"uller \& Kley 2012). We 
now study how planetesimal dynamics change in the case of a 
disk {\it uniformly} precessing at a constant rate $\dot 
\varpi_d$. 
Figure \ref{fig:ev_traj}b displays evolution of ${\bf e}_p$
for the same parameters as in panel (a) of that Figure, but 
in a disk precessing at the rate $\dot\varpi_d=A$. One can 
see that the main difference 
compared to the non-precessing case is that in the long run 
${\bf e}_p$ converged to the {\it limit cycle} behavior 
(Beaug\'e \etal 2010) rather than to a fixed point, as in 
panel (a). The sizes and shapes of the asymptotic limit cycles 
depend on both the planetesimal size $d_p$ and the disk precession 
rate $\dot \varpi_d$, as  discussed in detail in Appendix 
\ref{sect:quad_t_sol} and shown in Figure \ref{fig:limit_cycles}. 
This certainly complicates planetesimal dynamics.

To gain additional insights, in Appendix \ref{sect:lin_t_sol} 
we derive a full time-dependent solution for ${\bf e}_p$ in 
a precessing disk for the case of {\it linear} gas drag, when 
$\tau_d$ is independent of the relative particle-gas 
eccentricity $e_r$. This solution fully accounts for the 
gravitational and gas drag effects of the precessing disk as 
well as for the gravity of the binary companion. 

We use this solution as a basis for understanding planetesimal 
dynamics in a precessing disk in the more complicated but realistic 
case of quadratic gas drag. This regime, which does not admit 
general analytical solution even for the long-term behavior 
is explored in Appendix \ref{sect:quad_t_sol}. There we show 
that planetesimal dynamics with drag law (\ref{eq:F_drag}) 
depend on the relative role played by the binary companion, 
as described next.


\subsection{Strong binary perturbation case.}  
\label{sect:precess_bin}

Results of Appendices \ref{sect:lin_t_sol} \& \ref{sect:quad_t_sol} 
show that whenever binary gravity dominates ${\bf e}_p$ excitation 
and the condition
\ba
\left|\left(A-\dot\varpi_d\right)e_g+B_d\right|\lesssim |B_b|
\label{eq:bin_dom_cond}
\ea
is fulfilled, planetesimal dynamics proceed as if the disk
were not precessing: neither the gas eccentricity $e_g$ nor the
eccentricity driven by disk gravity $e_d$, equation 
(\ref{eq:e_forced_d}), are significant compared to the forced 
eccentricity due to binary $e_b=B_b/A$ (note that both binary 
and disk gravity contribute to $A$). 

In this case ${\bf e}_p$ is close to the relative 
planetesimal-gas eccentricity ${\bf e}_r$ and is 
approximately constant. As a result, planetesimal orbit 
maintains roughly fixed orientation with respect to the 
binary orbit and
\ba
k_p\approx k_b\frac{\left(A\tau_d\right)^2}{1+\left(A\tau_d\right)^2},
~~~
h_p\approx -k_b\frac{\left(A\tau_d\right)}{1+\left(A\tau_d\right)^2},
\label{eq:strong_bin}
\ea 
with $k_b$ defined by equation (\ref{eq:e_forced_b}).
Planetesimal orbits are aligned with the binary ($\varpi_p\to 0$) 
for $|A\tau_d|\to \infty$ (weak drag), but in the case of strong drag 
$|A\tau_d|\to 0$ planetesimal apsidal line points at 
$\varpi_d=270^\circ$, which agrees with Marzari \& Scholl (2000)
despite the disk precession. 

Interestingly, even though gas eccentricity $e_g$ does not appear in 
these expressions  (and neither does the precession rate 
$\dot\varpi_d$, at the lowest order) the effect of the gas drag is 
explicitly present via the non-trivial $\tau_d$ dependence. 
Thus, our precessing disk results obtained in the limit 
(\ref{eq:bin_dom_cond}) apply equally well to 
planetesimal dynamics in a purely axisymmetric ($e_g=0$) 
gaseous disk, extending the results of R13 to the case of 
non-zero gas drag --- note that $A$
in equation (\ref{eq:strong_bin}) and in the definition of $k_b$
is the full precession rate due to both binary and the disk.   

The value of $e_r$ in the regime (\ref{eq:bin_dom_cond}) is 
given by equations (\ref{eq:e_r}) and (\ref{eq:e_r_d}) with 
$d_c$ and $|A\tau_d|$ computed using $e_c\approx e_b =|B_b/A|$ 
(i.e. equation (\ref{eq:e_c}) in the limit $B_d\to 0$, 
$e_g\to 0$), see equations (\ref{eq:d_1}) and (\ref{eq:Atd}).


\subsection{Weak binary perturbation case.}  
\label{sect:precess_disk}

In the opposite case of weak driving of ${\bf e}_p$ by the 
binary companion we combine solutions (\ref{eq:steady}) 
and find the relative particle-gas eccentricity to be
\ba
e_r=\left|e_c^{\rm pr}\right|\frac{\left|A-\dot\varpi_d\right|\tau_d}
{\sqrt{1+\left(A-\dot\varpi_d\right)^2\tau_d^2}},
\label{eq:e_r_pr}
\ea
replacing equation (\ref{eq:e_r}) in the case of precessing disk.
Here we defined characteristic eccentricity 
\ba
e_c^{\rm pr}=-\frac{B_d}{A-\dot\varpi_d}-e_g,
\label{eq:e_prec}
\ea
which, according to SR13, is the relative particle-gas 
forced eccentricity in the no drag ($\tau_d\to \infty$) and 
no binary ($B_b\to 0$) case. As $\dot\varpi_d\to 0$ one finds 
$\left|e_c^{\rm pr}\right|\to e_c$ given 
by equation (\ref{eq:e_c}) with $k_b=h_b=0$; also, equation 
(\ref{eq:e_r_pr}) reduces to the non-precessing disk result 
(\ref{eq:e_r}). 

Plugging this expression for $e_r$ into equation (\ref{eq:tau_d})
one finds
\ba
\left|A-\dot\varpi_d\right|\tau_d=
\frac{d_p}{d_c^{\rm pr}}\left[\frac{1}{2}+
\sqrt{\frac{1}{4}+\left(\frac{d_c^{\rm pr}}{d_p}\right)^2}\right]^{1/2},
\label{eq:Atd_pr}
\ea
with a new characteristic planetesimal size 
\ba
d_c^{\rm pr} \equiv \frac{3C_D\mbox{E}
\left(\sqrt{3}/2\right)}{4\pi}\frac{n_p}
{\left|A-\dot\varpi_d\right|}\frac{\Sigma_g}{\rho_p}
\frac{r}{h}\left|e_c^{\rm pr}\right|.
\label{eq:d_1_pr}
\ea
These expressions are different from equations (\ref{eq:d_1}) and 
(\ref{eq:Atd}) in using $A-\dot\varpi_d$ instead of $A$ and 
$\left|e_c^{\rm pr}\right|$ instead of $e_c$. It is then clear that 
whenever a precessing disk dominates planetesimal dynamics equation 
(\ref{eq:e_r_d}) also holds provided that we replace 
$d_c\to d_c^{\rm pr}$ and $e_c\to \left|e_c^{\rm pr}\right|$.  
The same is true for our asymptotic results on ${\bf e}_p$ 
behavior presented in 
\S \ref{sect:strong}-\ref{sect:weak} if we also take $k_b\to 0$.

In the limit $\dot\varpi_d\to 0$ the value of ${\bf e}_f$ reduces 
to ${\bf e}_{f,d}$ given by equation (\ref{eq:e_f_d}). But when 
$|\dot\varpi_d|\gg |A|$ rapid disk precession suppresses 
excitation of planetesimal eccentricity by the disk gravity, 
i.e. the first term in equation (\ref{eq:e_prec}). 

It it worth noting that results of Appendix \ref{sect:lin_t_sol} 
for the case of linear drag suggest that neglecting binary gravity 
in the case of precessing disk might require a condition different 
from the direct opposite to the constraint (\ref{eq:bin_dom_cond}). 
Indeed, asymptotic solution (\ref{eq:e_r_weak}) for the relative 
eccentricity of planetesimals in the case of weak drag 
($\tau_{d,1},\tau_{d,2}\gg |A-\dot\varpi_d|^{-1}$) shows that the
term proportional to $k_b$ can be neglected only when 
\ba
\left|\left(A-\dot\varpi_d\right)e_g+B_d\right|\gtrsim |B_b|
\left(\frac{A-\dot\varpi_d}{A}\right)^2,
\label{eq:disk_dom_cond}
\ea
which is a more stringent criterion whenever 
$|\dot \varpi_d|\gg |A|$. The same constraint may be needed 
in the case of quadratic drag.  
However, in practice one often 
finds $|\dot\varpi_d|\lesssim |A|$, see Paper II in which case 
equation (\ref{eq:disk_dom_cond}) is just the 
opposite of the condition (\ref{eq:bin_dom_cond}).


\section{Diversity of planetesimal dynamics.}  
\label{sect:pl_dyn}


Results of \S \ref{sect:ecc_evol} demonstrate that the 
steady state value of the eccentricity vector ${\bf e}_p$ is 
fully determined by just two key parameters --- characteristic 
eccentricity $e_c$ and critical planetesimal size $d_c$, see equation 
(\ref{eq:e_r_d}). Eccentricity $e_c$ sets the overall scale of 
the ${\bf e}_p$, while $d_c$ is the planetesimal 
size at which planetesimal coupling to gas changes from weak to 
strong. We now explore the behavior of these 
variables as a function of system parameters to elucidate some 
important features of planetesimal dynamics.


\subsection{Behavior of $e_c$.}  
\label{sect:e_c}

\begin{figure}
\vspace{-.5cm}
\epsscale{1.25}
\plotone{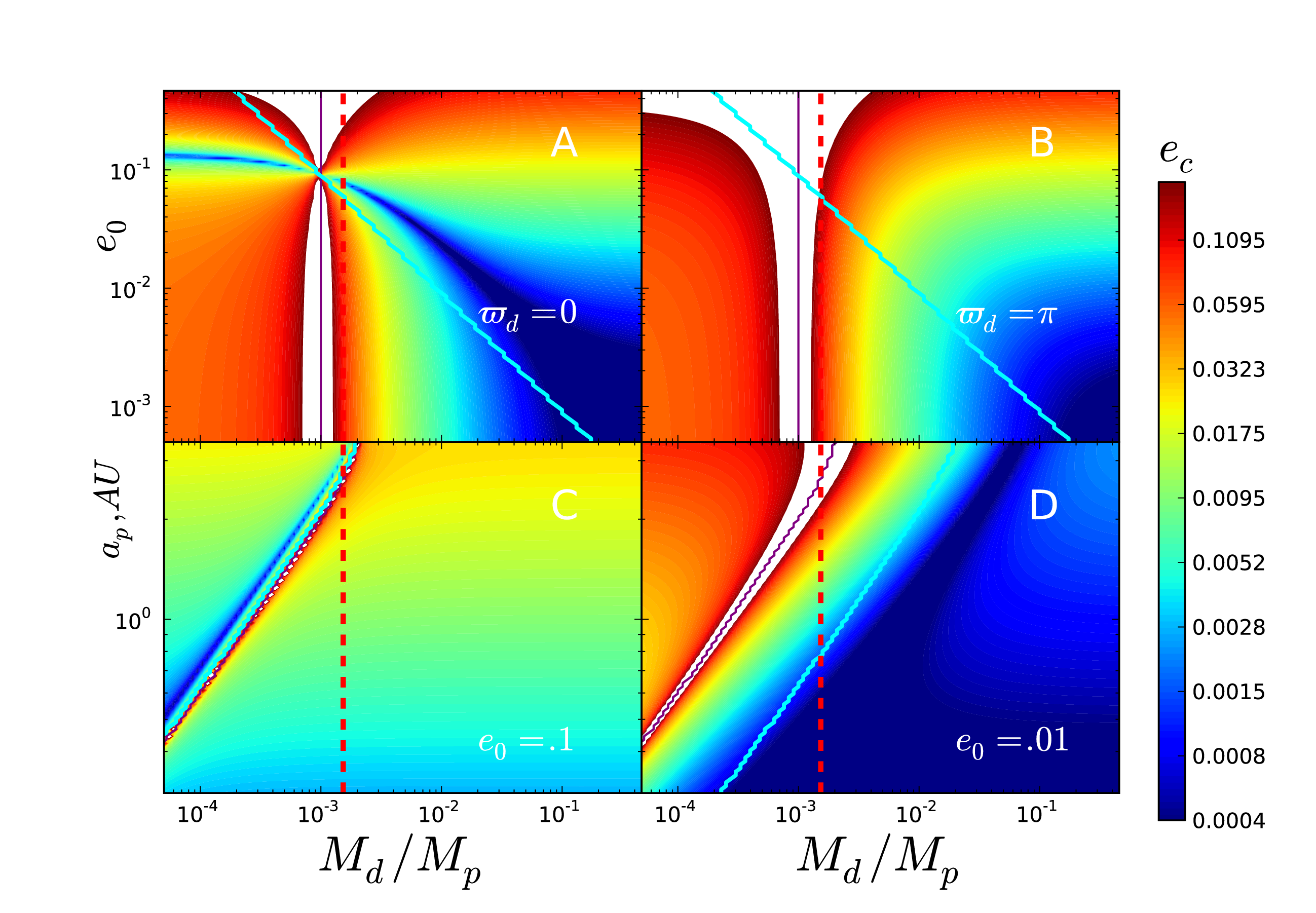}
\caption{
Map of the characteristic eccentricity $e_c$ as a function of 
$e_0$ and $M_d$ (upper panels) for two different 
disk orientations --- $\varpi_d=0$ (a) and (b) --- and as a 
function of $a_p$ and $M_d$ for two values of disk eccentricity 
$e_0$ at $a_{\rm out}$ (lower panels). Calculation is done 
for $\gamma$ Cep system at $a_p=2$ AU (the observed semi-major 
axis of the planet).  The dashed red line corresponds to 
$M_p \sin{i}$ for the observed planet in the $\gamma$ Cephei 
system.  The purple line is where $|A_d| = |A_b|$, and the 
blue line is where $|B_d| = |B_b|$. 
See text for details. }
\label{fig:e_c_map}
\end{figure}

In Figure \ref{fig:e_c_map}a,b we show $e_c$ computed for 
$\gamma$ Cep system at 2 AU --- the semi-major axis of its planet ---
as a function of disk mass $M_d$ and eccentricity $e_0$, for 
two disk orientations --- aligned ($\varpi_d=0$) and 
anti-aligned ($\varpi_d=\pi$) with the 
apsidal line of the binary.  

One can immediately see a feature 
common to both panels --- a narrow valley of high $e_c$ (white 
because of saturation at high $e_c$) at almost constant $M_d$. 
It appears because at this value of disk mass $A_d=-A_b$ and 
$A=0$, giving rise to a {\it secular} resonance. According to 
equations (\ref{eq:e_n_drag_decompose})-(\ref{eq:e_forced_d}) 
and (\ref{eq:e_c}) $e_c$ gets driven to 
high values as $A\to 0$. This resonance has been previously  
discussed in R13 and SR13. 

Equations (\ref{eq:A_b})-(\ref{eq:A_d}) predict that at a 
given distance from the primary $a_p$ this resonance occurs 
for the disk mass
\ba
M_{d,A=0} &=& M_s\frac{3}{4(2-p)|\psi_1|}
\left(\frac{a_p}{a_b}\right)^{1+p}
\left(\frac{a_{\rm out}}{a_b}\right)^{2-p}
\label{eq:Md_A_0}\\
& \approx & 1.5\times 10^{-3}M_\odot \frac{M_s}{0.4~M_\odot}
\frac{a_{\rm out,5}}{a_{b,20}^3}a_{p,2}^2, 
\nonumber
\ea
where $a_{\rm out,5}\equiv a_{\rm out}/(5$ AU), 
$a_{p,2}\equiv a_p/(2$ AU), and $a_{b,20}\equiv a_b/(20$ AU).
This estimate agrees with Figure \ref{fig:e_c_map}a,b for the 
$\gamma$ Cep parameters and a disk with $p=1$ and $\psi_1(p=1)=-0.5$ 
(SR13).

Existence of this resonance is independent of the 
relative disk-binary orientation because planetesimal 
precession rates $A_b$ and $A_d$ are determined by the 
{\it axisymmetric} components of the binary and disk 
gravitational potentials. For this reason $M_{d,A=0}$ is 
the same for all disk orientations. 
To the right of the secular resonance disk gravity 
dominates planetesimal precession rate and suppresses $e_c$ if 
disk eccentricity is small (R13). 

At high disk eccentricity, typically $e_0\gtrsim 0.05$, this 
suppression vanishes because for large $M_d\gtrsim 10^{-3}M_\odot$ 
disk gravity starts to dominate ${\bf e}_p$ excitation. This 
statement is true above the blue line $|B_b|=|B_d|$ in 
Figure \ref{fig:e_c_map}a,b (the origin of the low-$e_c$ band at
small $M_d$ and high $e_0$ in Figure \ref{fig:e_c_map}a is 
discussed in \S \ref{sect:valley}). Further increase of the disk 
mass in this region does not affect $e_c$ because planetesimal 
dynamics switches to the so-called DD regime (SR13) in which 
$e_p(a_p)\approx |\psi_2/\psi_1|e_g(a_p)$, independent of $M_d$. 
As a result, high $e_g$ leads to high $e_p$.

 In Figure \ref{fig:e_c_map}c,d we explore the dependence of $e_c$ 
on the distance from the binary $a_p$ and $M_d$ for two different 
values of the disk eccentricity $e_0=0.1$ and 0.01. Here we look 
only at an aligned disk case. Again, an obvious feature of 
these maps is the secular resonance around the blue dashed 
curve for $|A_d|=|A_b|$, where $e_c$ is very large and 
collisional growth is impossible. In Figure \ref{fig:e_c_map}d
there is also a ``valley'' of low $e_c$ to the right from the 
blue line $|B_b|=|B_d|$, whose origin is discussed in \S 
\ref{sect:valley}. 

These maps make it clear that $e_c$ becomes independent 
of $M_d$ (at a given separation $a_p$) when the disk mass becomes 
large enough. This is a direct consequence of the planetesimal 
dynamics switching into the DD regime (SR13), when both 
eccentricity excitation and apsidal precession of planetesimals 
are dominated by the disk gravity with negligible contribution 
from the binary companion. In the high-$M_d$ regime $e_c$ 
decreases as $a_p$ goes down. This is a consequence of our 
adopted disk model, in which $e_d\propto a_p$ and the fact that 
$e_c\propto e_d$ in the DD regime.


\subsection{Valley of stability in aligned disks.}  
\label{sect:valley}

Figure \ref{fig:e_c_map}a,b shows that irrespective of the 
disk orientation $e_c$ is low for high $M_d\gtrsim 10^{-2}M_\odot$
and small disk eccentricity, $e_0\lesssim 10^{-2}$. Outside  
this corner of phase space $e_c$ is much higher, which makes
planetesimal growth problematic there. At the same time, in the 
case of an aligned disk ($\varpi_d=0$) low values of $e_c$ are
also possible in a narrow ``valley'' stretching towards high 
$e_0$ and low $M_d$. Since this feature may have interesting 
implications for planet formation in binaries (see Paper II for 
details) we discuss its origin in more detail.

Equation (\ref{eq:e_c}) implies that in an aligned disk 
$h_g=h_d=0$ so that
\ba
e_c\approx |k_g-k_d-k_b|=\left|\frac{B_b+B_d+A e_g}{A}\right|.
\label{eq:e_c_aligned}
\ea
For massive disks, to the right from the vertical 
$|A_b|=|A_d|$ line in Figure \ref{fig:e_c_map}a, one can 
set $A\approx A_d$ and relate it 
to $B_d$ via equations (\ref{eq:A_d}) and (\ref{eq:B_d}). As a 
result, equation (\ref{eq:e_c_aligned}) becomes 
\ba
e_c\approx \left|\frac{B_b+B_d\left(1+2\psi_1\psi_2^{-1}
\right)}{A_d}\right|.
\label{eq:e_c_aligned1}
\ea
For the disk model considered here ($p=1$, $q=-1$) one has 
$\psi_1=-0.5$, $\psi_2=1.5$ and $1+2\psi_1\psi_2^{-1}=1/3$ so that
$e_c\approx |A_d|^{-1}\left|B_b+B_d/3\right|$. 
Also $B_d>0$ while $B_b$ is always negative, see equations 
(\ref{eq:B_b})-(\ref{eq:B_d}). Given that $B_d\propto e_0 M_d$
it is then obvious that one can make $e_c\approx 0$ by  
choosing $e_0 M_d$ such that $|B_b|\approx |B_d|/3$. 
Thus, in the case of an aligned disk 
a ``valley'' of low $e_c$ is described by the relation 
$e_0\propto M_d^{-1}$ as long as $|A_d|\gtrsim |A_b|$ 
(i.e. for massive disks). 

From this discussion we see that $e_c\approx 0$ for values 
of $e_0$ and $M_d$, which are close to the curve
\ba
M_{d,|B_b|=|B_d|} &=& M_s\frac{15}{8(2-p)|\psi_2|}
\frac{e_b}{e_0}\\
\nonumber
&\times & \left(\frac{a_p}{a_b}\right)^{2+p+q}
\left(\frac{a_{\rm out}}{a_b}\right)^{2-p-q}
\label{eq:Md_B_B}\\
& \approx & 1.2\times 10^{-3}M_\odot \frac{M_s}{0.4~M_\odot}
\frac{e_b}{0.4}\frac{0.1}{e_0}
\frac{a_{\rm out,5}^2}{a_{b,20}^4}a_{p,2}^2, 
\nonumber
\ea
on which $|B_b|=|B_d|$, see equations (\ref{eq:B_b})-(\ref{eq:B_d})
in which we took $p=1$, $q=-1$.
This relation is shown by the blue line in Figure 
\ref{fig:e_c_map} and is quite close to the valley of 
low $e_c$.

Note that according to equation (\ref{eq:e_c_aligned1}) the 
value of $e_c$ can be lowered {\it globally} in a massive disk 
if its structure is such that $1+2\psi_1\psi_2^{-1}=0$. 
However, this is not the case for the disk model used in 
this work.

The situation is different for the low mass, aligned disks
to the left of the $|A_d|= |A_b|$ (blue dashed) line in 
Figure \ref{fig:e_c_map}a. Here $A\approx A_b$ and $B_b$ 
dominates over $B_d$ for low enough $M_d$ at a fixed $e_0$,
which in terminology of SR13 corresponds to the Case BB
of planetesimal excitation. In this regime equation 
(\ref{eq:e_c_aligned}) shows that 
\ba
e_c\to\left|e_g+\frac{B_b}{A_b}\right|=
\left|e_g-\frac{5}{4}\frac{a_p}{a_b}e_b\right|
\label{eq:BB}
\ea
Our adopted radial scaling of $e_g$ in the form (\ref{eq:Sigma0}) 
with $q=-1$ results in a particular value of 
\ba
e_0\big|_{e_c\to 0}=\frac{5}{4}\frac{a_{\rm out}}{a_b}e_b=
0.125\frac{a_{\rm out}/a_b}{0.25}\frac{e_b}{0.4},
\label{eq:e_0_crit}
\ea
for which $e_c\to 0$. This critical value 
of $e_0$ in independent of $M_d$ explaining why the valley
of low $e_c$ starts going almost horizontally for 
$M_d\lesssim M_d\big|_{A=0}$ in Figure 
\ref{fig:e_c_map}a.

Moreover, $e_0|_{e_c\to 0}$ is also independent of $a_p$, 
which means that $e_c\to 0$ {\it globally} when
$e_0\to e_0|_{e_c\to 0}$ in parts of the disk where 
$|A_b|\gtrsim |A_d|$ and $|B_b|\gtrsim |B_d|$. This is the 
reason why in the upper left corner of Figure 
\ref{fig:e_c_map}c $e_c$ is considerably lower than in the 
same region of Figure \ref{fig:e_c_map}d, despite $e_0$
being an order of magnitude higher in the former case.
Indeed, according to equation (\ref{eq:e_0_crit}) $e_0=0.1$ 
used in Figure \ref{fig:e_c_map}c is very close to 
$e_0|_{e_c\to 0}$ for the adopted system parameters. As a 
result of this coincidence, $e_c$ is strongly suppressed in
the BB regime in a rather eccentric ($e_0=0.1$) disk.

A narrow region of low $e_c$ stretching along the blue 
curve $|B_b|=|B_d|$ in Figure \ref{fig:e_c_map}c,d is 
the same valley of stability, but now revealing itself 
in $M_d-a_p$ coordinates\footnote{Curves of 
$|A_d|=|A_b|$ and $|B_d|=|B_b|$ run parallel to each other in 
Figure \ref{fig:e_c_map}c,d because $e_d\propto a_p$ in our 
disk model, see equations (\ref{eq:A_b})-(\ref{eq:B_d}).} . 
It may lie inside (for low $e_0$) as well as outside (for 
high $e_0$) of the secular resonance. Note that in Figure 
\ref{fig:e_c_map}c the $|A_d|=|A_b|$ and $|B_d|=|B_b|$ 
curves fall almost on top of each other, which is a 
coincidence caused by our choice of $e_0=0.1$ in this case. 
Because of that the valley of stability appears as a very 
narrow band of low $e_c$ just to the left of the $|B_d|=|B_b|$ 
curve in this panel. 

If the disk is not aligned with the binary orbit and $\varpi_d$ is
not small then both $h_d$ and $h_g$ are nonzero and contribute 
to $e_c$, see equation (\ref{eq:e_c}). Moreover, for disks which 
are close to being anti-aligned with the binary, $k_b$ and $k_d$
have the same sign, eliminating 
the possibility of their mutual cancellation. As a result,
the low-$e_c$ valley at high $e_0$ and low $M_d$ disappears as long 
as $|\varpi_d-\varpi_b|\gtrsim 10^\circ$.

To summarize, the valley of stability creates favorable 
conditions for lowering planetesimal velocity in aligned disks 
{\it locally}, around some particular locations, even in low 
mass disks with $M_d\lesssim 10^{-2}M_\odot$. 

\begin{figure}
\epsscale{1.2}
\plotone{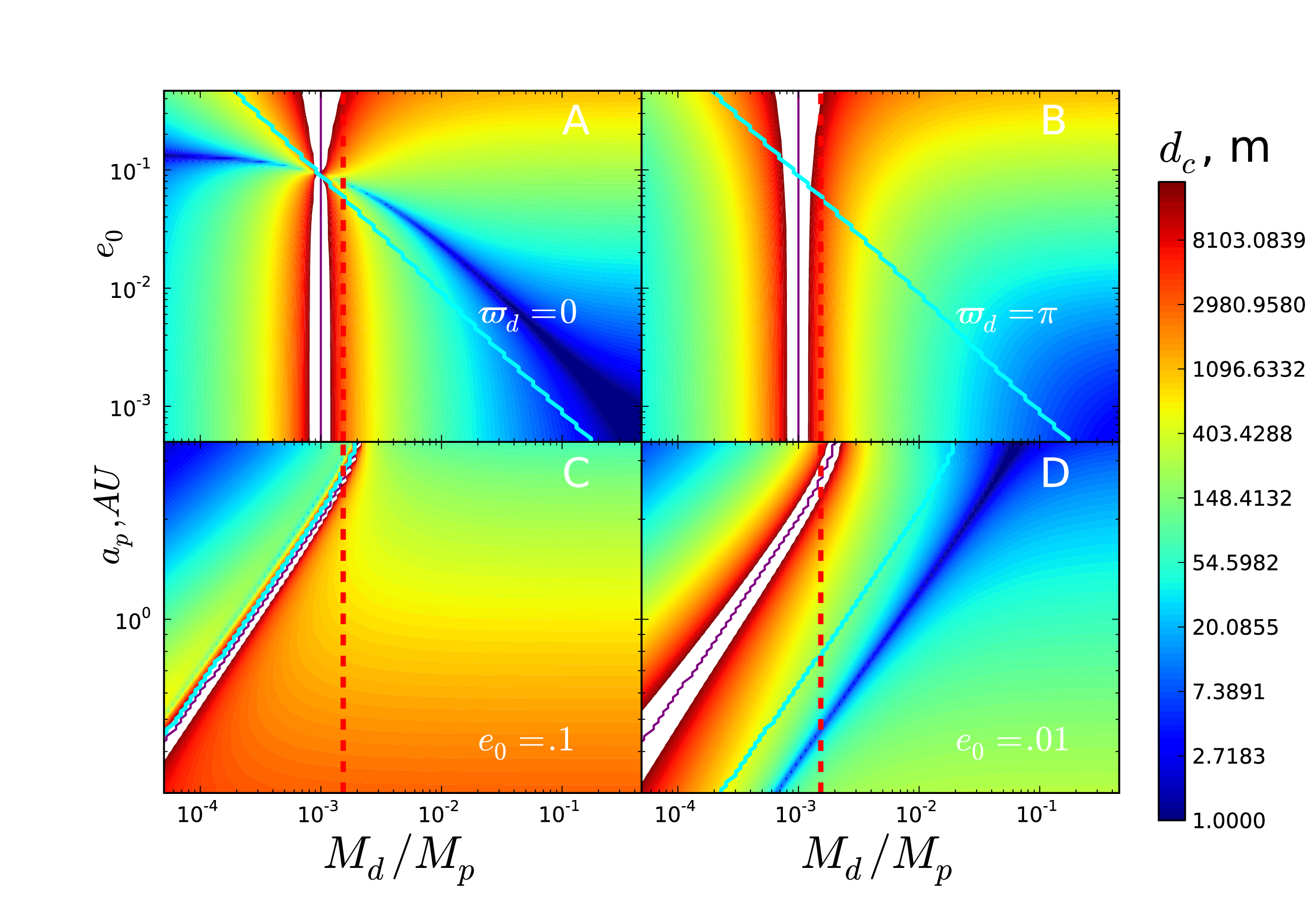}
\caption{
Same as Fig. \ref{fig:e_c_map} but for the behavior of the 
characteristic size $d_c$ given by equation (\ref{eq:d_1}).
\label{fig:d_c_map}}
\end{figure}


\subsection{Behavior of $d_c$.}  
\label{sect:d_c}

Next we discuss the behavior of the characteristic size $d_c$
at which planetesimals of similar (but not equal) mass collide 
at highest relative velocity $\sim e_cv_K$. 
Equation (\ref{eq:d_1}) makes it clear that {\it for a given 
value of} $e_c$ critical size is a sensitive function of the 
planetesimal precession rate $A$: $d_c$ is smaller for higher 
$|A|$. If planetesimal precession is dominated by the potential 
of the secondary then $A=A_b$, and one finds 
\ba
d_c &=& \frac{C_D {\rm E}(\sqrt{3}/2)}{\pi\nu}\frac{r}{h}
\frac{\Sigma_g}{\rho_p}\left(\frac{a_b}{a_p}\right)^3e_c
\label{eq:d_c_b}
\\
 &\approx & 30~\mbox{km}~\frac{C_D}{\nu}\frac{0.1}{h/r}
\frac{M_{d,-2}a_{b,20}^3}{a_{\rm out,5}}\frac{e_c}{0.1}a_{p,1}^{-4}
\nonumber
\ea
where the numerical estimate is for $p=1$ disk and $\rho_p=3$
g cm$^{-3}$. 

In the opposite case, when precession is dominated by the 
disk gravity and $A=A_d$ one obtains
\ba
d_c &=& \frac{3C_D {\rm E}(\sqrt{3}/2)}{8\pi^2\psi_1}\frac{r}{h}
\frac{M_p}{\rho_p a_p^2}e_c
\label{eq:d_c_d}
\\
 &\approx & 1~\mbox{km}~\frac{C_D}{\psi_1}\frac{0.1}{h/r}
\frac{e_c}{0.1}M_{p,1}a_{p,1}^{-2},
\nonumber
\ea
independent of the disk mass. It is obvious that in the 
disk-dominated case $d_c$ is much smaller than in the 
binary-dominated case for $a_p\lesssim 1$AU, a fact  
predicted in R13. 

{\ This difference 
can be easily seen in Figure \ref{fig:dif_dyn_approx}, where
the situation depicted in panel (a) corresponds to the DD 
regime, in which equation (\ref{eq:d_c_d}) applies. As a result,
the planetesimal size for which the low-$e_r$ ``waist'' in this 
Figure is narrowest is around 1 km. On the contrary, 
Figure \ref{fig:dif_dyn_approx}b shows a situation in which 
disk gravity has been turned off, so the dynamics are in
the BB regime and equation (\ref{eq:d_c_b}) applies. Not
surprisingly, this pushes the characteristic $d_p$ at the
narrowest point of the waist to be about 30 km. 

Using this reasoning one might expect the critical ``dangerous'' 
size $d_p$ at which $e_r\sim e_c$ for objects of comparable size 
to be smaller for more massive disks in which $|A_d|\gg |A_b|$. 
However, this logic directly applies only if $e_c$ were kept 
the same. In reality, changing $A$ also directly affects the 
value of $e_c$, see equation (\ref{eq:e_c}). Figure 
\ref{fig:d_c_map} shows that in practice the behavior of 
$d_c$ largely reflects that of $e_c$, with all the features 
of $e_c$ maps (e.g. valleys of low $d_c$) present in $d_c$ 
maps as well. In particular, the valley of stability shows 
up prominently in Figure \ref{fig:d_c_map}a,d.

The only noticeable difference with Figure \ref{fig:e_c_map} 
is the {\it increase} of $d_c$ with decreasing $a_p$ in the 
high-$M_d$ (DD) regime, see Figure \ref{fig:d_c_map}c,d, a 
behavior which is predicted 
by equation (\ref{eq:d_c_d}). Also, in agreement with equation 
(\ref{eq:d_c_b}), $d_c$ decreases with increasing $a_p$ in the 
outer disk for small $M_d$ (upper left in Figure \ref{fig:d_c_map}c,d)
even though $e_c$ varies there weakly. In this region planetesimal 
dynamics is determined predominantly by the binary companion 
(BB regime of SR12) and equation (\ref{eq:d_c_b}) applies.


\section{Distribution of relative planetesimal velocities.}  
\label{sect:rel_vel}


Our next step is to study the behavior of the relative 
approach velocity $v_{12}$ between planetesimals with sizes $d_1$ 
and $d_2$. It is this velocity that determines the outcome of 
their collision. 

We now provide a calculation of the 
{\it distribution $df_{12}/dv_{12}$ of $v_{12}$} between the two 
planetesimal populations, one with eccentricity vector 
${\bf e}_p(d_1)$ and another with ${\bf e}_p(d_2)$. 
In previous sections we have shown that after the initial 
transient period when the free eccentricity damps out, 
the value of ${\bf e}_p$ becomes time-independent and is 
uniquely determined by the planetesimal size. Then the only 
additional orbital parameter that can give rise to the variation 
of the relative velocity $v_{12}$ is the difference in semi-major 
axes $b_{12}$ between approaching particles, see equation 
(\ref{eq:v_unper}) of Appendix \ref{sect:local}. Using 
equations (\ref{eq:rel_unpertxy}), (\ref{eq:v_unper}) 
it can be written as
\ba
v_{12}=\Omega a_p\left[e_{12}^2-\frac{3}{4}
\left(\frac{b_{12}}{a_p}\right)^2\right]^{1/2},
\label{eq:v_unper_b}
\ea
where $a$ is the mean semi-major axis of both planetesimals, 
and the condition of close approach $x_{12}=0$ was used. Note 
that in this expression we ignored the contribution of 
particle inclination to the velocity. This is a reasonable 
assumption since we expect eccentricity excitation in the 
binary plane to dominate over the out of plane excitation.

\begin{figure}
\epsscale{1.2}
\plotone{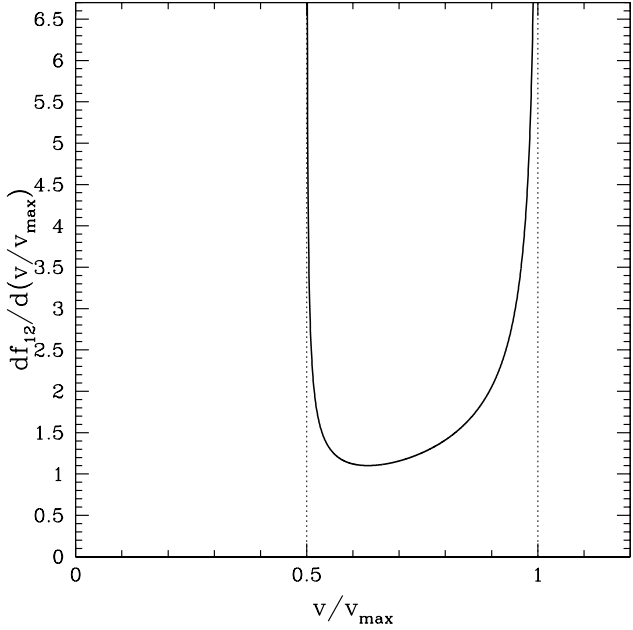}
\caption{
Distribution of the relative approach velocity $v_{12}$ of 
colliding planetesimals given by equation (\ref{eq:dfdv2}). 
Relative velocity is normalized by its maximum value 
$v_{max}=e_{12} n_p r$, where $e_{12}$ is the relative 
eccentricity of the two planetesimals, which is a function 
of their sizes, see \S 
\ref{sect:strong_strong}-\ref{sect:weak_strong}. Minimum 
approach velocity is $v_{max}/2$. 
\label{fig:vel_dist}}
\end{figure}

Ida \etal (1993) consider encounters between the two populations of 
objects with fixed eccentricity vectors ${\bf e}_1=(k_1,h_1)$ and 
${\bf e}_2=(k_2,h_2)$. They derive the following expression for the flux of 
objects with eccentricity ${\bf e}_2$ approaching a given object 
with eccentricity ${\bf e}_1$ with random orbital phases, having 
separation of their semi-major axes $b_{12}$ in the range 
$(b_{12},b_{12}+db_{12})$:
\ba
dF_{12}=\frac{1}{\pi^2}\frac{\Sigma_2}{a_p m_2 i_{12}}\frac{v_{12}db_{12}}
{\left[e_{12}^2 -(b_{12}/a_p)^2\right]^{1/2}}.
\label{eq:flux_raw}
\ea
Here $e_{12}=\left[(h_1-h_2)^2+(k_1-k_2)^2\right]^{1/2}$ is the 
relative eccentricity between the two particle populations, $i_{12}$
is their relative inclination, and 
$\Sigma_2$ is the surface density of objects with eccentricity 
${\bf e}_2$. 

Using equation (\ref{eq:v_unper_b}) we can express $db$
in equation (\ref{eq:flux_raw}) via $dv_{12}$, resulting in
differential particle flux per unit $v_{12}$
\ba
\frac{dF_{12}}{dv_r}=\frac{4}{3\pi^2}\frac{\Sigma_2 a_p}{m_2 i_{12}}
\frac{e_{12}^2 -(3/4)(b_{12}/a_p)^2}
{|b_{12}|\left[e_{12}^2 -(b_{12}/a_p)^2\right]^{1/2}}.
\label{eq:dfdv}
\ea
We now express $b_{12}$ via $v_{12}$ using equation (\ref{eq:v_unper_b})
and introduce 
\ba
v_{\rm min}=\frac{1}{2}e_{12} n a_p, ~~~ v_{\rm max}=e_{12} n a_p.
\label{eq:v_ext}
\ea 
Then it is clear that $v_{\rm min}<v_{12}<v_{\rm max}$ and we can 
re-write (\ref{eq:dfdv}) as
\ba
\frac{dF_{12}}{dv_{12}}=\frac{1}{\pi^2}\frac{\Sigma_2}{m_2 i_{12}}
\frac{v_{12}^2}
{\left[\left(v_{\rm max}^2-v_{12}^2\right)
\left(v_{12}^2-v_{\rm min}^2\right)\right]^{1/2}}.
\label{eq:dfdv1}
\ea

From this we find that the distribution of relative velocities 
$df_{12}/dv_{12}$ of different planetesimals normalized to unity is given 
by the following expression:
\ba
\frac{df_{12}}{dv_{12}}=\frac{v_{\rm max}^{-1}}{\mbox{E}\left(\sqrt{3}/2\right)}
\frac{v_{12}^2}
{\left[\left(v_{\rm max}^2-v_{12}^2\right)
\left(v_{12}^2-v_{\rm min}^2\right)\right]^{1/2}}.
\label{eq:dfdv2}
\ea
Particle sizes enter into this expression only through $e_{12}$
via equations (\ref{eq:v_ext}). 

This distribution of relative velocities is shown in Figure 
\ref{fig:vel_dist}. It diverges at both $v=v_{\rm min}$ and 
$v=v_{\rm max}$, but the total particle flux is finite and given by
\ba
F_{12}=\int\limits_{v_{\rm min}}^{v_{\rm max}}\frac{dF_{12}}{dv_{12}}dv_{12}=
\frac{\mbox{E}\left(\sqrt{3}/2\right)}{\pi^2}
\frac{\Sigma_2 e_{\rm max}}{m_2 i_{12}}
\label{eq:flux_tot}
\ea
With distribution function (\ref{eq:dfdv1}) one finds the mean 
relative velocity $\langle v_{12}\rangle\approx 0.81v_{\rm max}=
0.81e_{12} n_p a_p$, while the rms velocity is given by 
$v_{\rm rms}=\langle v_{12}^2\rangle^{1/2}=0.828 e_{12} n_p a_p$.


\section{Relative velocity between planetesimals.}  
\label{sect:char_vel}


The results of the previous section clearly demonstrate that 
the relative velocity with which two planetesimals with 
sizes $d_1$ and $d_2$ approach each other prior to collision 
is determined by their relative eccentricity 
$e_{12}=|{\bf e}_p(d_1)-{\bf e}_p(d_2)|$. Using solutions 
(\ref{eq:k_p})-(\ref{eq:h_p}) it is trivial to show that 
\ba
e_{12}=e_c\frac{\left|A\tau_{d,1}-A\tau_{d,2}\right|}
{\sqrt{\left(1+A^2\tau_{d,1}^2\right)
\left(1+A^2\tau_{d,2}^2\right)}},
\label{eq:erel}
\ea
where $\tau_{d,i}\equiv \tau_{d}(d_i)$, $i=1,2$.
According to the results of \S \ref{sect:ecc_evol}, $A\tau_d$
and, subsequently, $e_{12}$, are functions of (1) sizes of 
the colliding planetesimals $d_{1,2}$ and (2) binary parameters 
and local disk properties, which set the values of both 
$e_c$ and $d_c$, see equations (\ref{eq:e_c}) and (\ref{eq:d_1}). 
We already explored the latter in \S \ref{sect:pl_dyn} and now 
we turn our attention to understanding $e_{1,2}(d_1,d_2)$. 

In Figure \ref{fig:e_r_map} we map out $e_{12}(d_1,d_2)$ (as 
well as the relative velocity $v_{12}=e_{1,2}v_K$) at the location 
of the planet} $a_p=2$ AU in the $\gamma$ Cephei system for different 
characteristics of the disk, for which a model (\ref{eq:Sigma0}) 
with $p=1$, $q=-1$ is adopted. We vary disk mass $M_d$, eccentricity 
at its outer edge $e_0$, and its orientation with respect to the 
binary orbit $\varpi_d$, one at a time keeping other disk parameters 
fixed. All panels clearly show several key invariant features. 

First, there is a critical size of order $d_c$, around 
$d_1=d_2\sim (0.1-1)$ km, at which maps exhibit a ``waist'', 
in which $e_{12}$ is small for collisions of equal size 
bodies. Second, $e_{12}$ becomes small for encounters between 
both the small bodies, with $d_1,d_2\lesssim d_c$, and for large 
objects with $d_1,d_2\gtrsim d_c$. Third, $e_{12}$ saturates 
at a value roughly independent of $d_1$ or $d_2$ for collisions 
of particles with very different sizes, i.e. when 
$d_1\lesssim d_c\lesssim d_2$, and vice versa.

These gross features, as well as the variations of the overall 
velocity scale seen in these maps, are addressed below  
using the results of \S \ref{sect:ecc_evol}. Given that 
particles can be in different drag regimes --- strong or weak --- 
we will consider several possibilities separately.

\begin{figure}
\epsscale{1.2}
\plotone{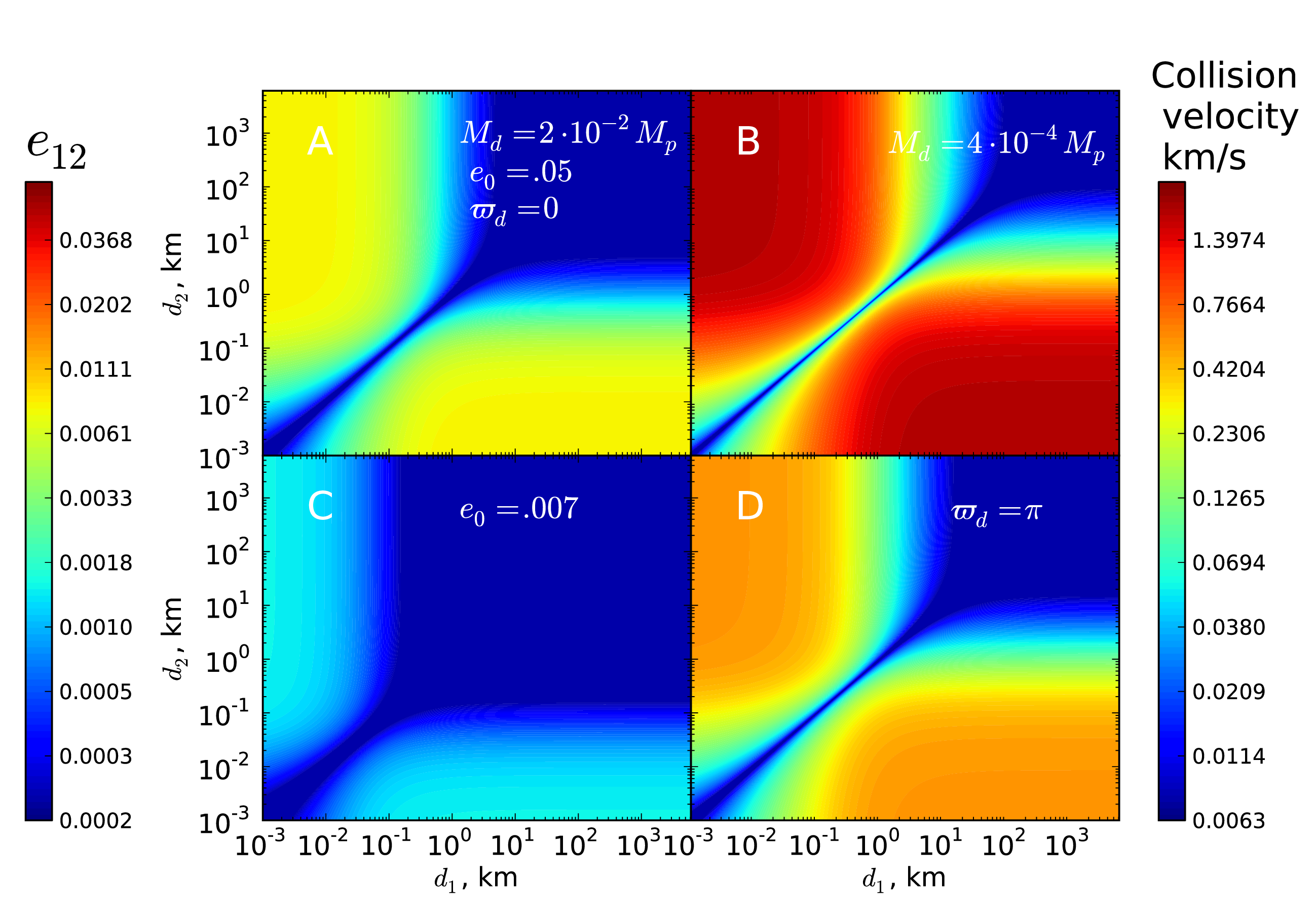}
\caption{
Relative approach velocity (right color bar) and relative 
eccentricity (left color bar) of planetesimals with sizes 
$d_1$ and $d_2$ experiencing close approach.
Calculation is done for $\gamma$ Cephei system at $2$ AU assuming 
an eccentric disk with $p=1$, $q=-1$ and other disk parameters --- 
$M_d$, $e_0$, $\varpi_d$ --- varying as indicated on the panels. 
Eccentricity and planetesimal size scales $e_c$ and $d_c$
in different panels can be inferred from Figures 
\ref{fig:e_c_map}a,b and \ref{fig:d_c_map}a,b.
\label{fig:e_r_map}}
\end{figure}


\subsection{Strong-strong encounters.}  
\label{sect:strong_strong}

When both planetesimals are in strong drag regime, 
$d_1,d_2\ll d_c$, both 
$\left|A\tau_{d,1}\right|\ll 1$ and  
$\left|A\tau_{d,2}\right|\ll 1$. Then equation (\ref{eq:erel})
predicts that
\ba
e_{12}^{\rm ss} & \approx & e_c\left|\left|A\tau_{d,1}\right|-
\left|A\tau_{d,2}\right|\right|
\label{eq:ss_er}\\
& \approx & e_c\left|\left(\frac{d_1}{d_c}\right)^{1/2}-
\left(\frac{d_2}{d_c}\right)^{1/2}\right|.
\label{eq:ss_er1}
\ea
where we used equation (\ref{eq:Atd_s}) to express 
$\left|A\tau_d\right|$ in terms of planetesimal sizes.
Since $d_{1,2}\ll d_c$ in the strong drag limit, one finds 
that $e_r^{\rm ss}\lesssim e_c$, which explains low values 
of $e_r$ in the lower left corner in maps in Figure 
\ref{fig:e_r_map}.

Physically, in this regime relative velocity of two planetesimals is 
considerably lower than their individual velocities because 
of the apsidal alignment of their orbits by gas drag, 
see Marzari \& Scholl (2000)) and similar magnitudes of 
${\bf e}_p$.


\subsection{Weak-weak encounters.}  
\label{sect:weak_weak}

When both planetesimals are in the weak drag regime
$\left|A\tau_{d,1}\right|\gg 1$ and  
$\left|A\tau_{d,2}\right|\gg 1$, 
one finds using equation (\ref{eq:erel}) that
\ba
e_{12}^{\rm ww} & \approx & e_c\left|\left|A\tau_{d,1}\right|^{-1}-
\left(A\tau_{d,2}\right|^{-1}\right|
\label{eq:ww_er}\\
& \approx & e_c\left|\frac{d_c}{d_1}-
\frac{d_c}{d_2}\right|.
\label{eq:ww_er1}
\ea
where equation (\ref{eq:Atd_w}) has been used. Since 
$d_{1,2}\gg d_c$ in the weak drag limit, one again finds 
that $e_r^{\rm ww}\lesssim e_c$, explaining the low relative 
eccentricity in the upper right corner in maps in Figure 
\ref{fig:e_r_map}.

In this case apsidal alignment is again at work, lowering 
$e_r$ compared to $e_p(d_1),e_p(d_2)$. However, now it 
is caused by the disk+binary gravity, which affects 
planetesimals in the same way when they are weakly 
coupled to gas. This is because the gas damps the free 
eccentricity, but is not strong enough to significantly 
change the forced eccentricity.


\subsection{Weak-strong encounters.}  
\label{sect:weak_strong}

When one of the planetesimals (e.g. of size $d_1$) is in the strong 
drag regime, $\left|A\tau_{d,1}\right|\ll 1$, while the other 
is in the weak drag regime, $\left|A\tau_{d,2}\right|\gg 1$, 
equation (\ref{eq:erel}) shows that their relative eccentricity 
$e_{12}$ is just 
\ba
e_{12}^{\rm sw}\approx e_c.
\label{eq:sw_er}
\ea
One can see that $e_{12}$ is roughly independent of the sizes 
of particles participating in an encounter.


\subsection{Overall $e_{12}$ scale as a function of disk parameters.}  
\label{sect:summ_rel_vel}

The overall scale of $e_{12}$ in each of the maps shown in 
Figure \ref{fig:e_r_map} is characterized by $e_{12}$ in one 
of the high-velocity corners. According to \S \ref{sect:weak_strong}
this scale is just $e_c$, which allows us to use the results of 
\S \ref{sect:pl_dyn} to understand how the typical $e_{12}$
varies as we change the disk parameters. Note that in 

Comparison of panels (a) and (b) of Figure \ref{fig:e_r_map}
shows that disk mass $M_d$ plays an important role in setting 
$e_{1,2}$: planetesimals in low mass disks
($M_d=4\times 10^{-4}M_\odot$) collide with much higher 
speeds than in higher mass ($M_d=2\times 10^{-2}M_\odot$) 
disk. This is because for the chosen value of $e_0=0.05$
the low mass disk is in the BB regime and the value of 
$e_c\approx 0.05$ is high, see Figure \ref{fig:e_c_map}a. 
Increasing $M_d$ as in panel (a) brings the disk in the 
DD regime and also close to the valley of stability. For 
that reason, in higher mass disk with $M_d=0.02M_\odot$ 
one gets much lower $e_c\approx 0.008$.  

Lowering $e_0$ for a high mass disk as in panel (c) 
reduces relative velocity scale even more, simply 
because for $e_0=0.007$ the system gets even deeper into
the valley of stability, where the corresponding  
$e_c\approx 1.5\times 10^{-3}$, see Figure \ref{fig:e_c_map}a.  

Comparison of panels (a) and (d) shows that changing disk 
orientation also strongly affects $e_r$: there is no valley 
of stability in the misaligned disk and characteristic 
eccentricity scale becomes $e_c\approx 0.014$. As a result, 
particles in a mis-aligned disk collide at higher 
speeds than in the aligned disk.


\section{Discussion.}  
\label{sect:disc}


Our work extends and complements existing results on 
planetesimal dynamics in binaries in several important ways. 

First, for the first time, our solutions for ${\bf e}_p$ in 
\S \ref{sect:ecc_evol} 
{\it simultaneously} account for a number of key physical ingredients 
needed for a complete description of secular dynamics of 
planetesimals in binaries: gravity of both eccentric
disk and eccentric companion as well as the gas drag, which causes 
orbital phasing of planetesimals and reduces their relative
eccentricity in certain regimes. 

Second, we provide a rigorous derivation of the 
equations of eccentricity evolution due to gas drag 
(\ref{eq:kdrag})-(\ref{eq:tau_d}) in an {\it eccentric}
disk. Previously, Adachi \etal (1976) derived analogous equations for
the case of a circular disk, while Beaug\'e \etal (2010) 
proposed a set of
empirical equations similar to (\ref{eq:kdrag})-(\ref{eq:hdrag}) 
but without proper calculation of the constant pre-factors.  

Third, we derive an analytic expression (\ref{eq:dfdv2}) for 
the relative velocity distribution function $df_{12}/dv_{12}$ 
for locally homogeneous populations of objects with fixed eccentricity 
vectors, which is appropriate in the limit $|{\bf e}_p|\ll 1$ 
in the presence of gas drag. We also provide an in-depth analysis
of $e_{12}$ behavior for objects of different sizes in 
systems with different parameters (\S \ref{sect:char_vel}). 
Previously the distribution of planetesimal encounter velocities 
has been explored only numerically, by following a large 
number of trace particles in simulations of different kinds 
(Th\'ebault \etal 2006, 2008, 2009; Paardekooper \etal 2008; 
Fragner \etal 2011). 
Thus, our derivation of $df_{12}/dv_r$ represents an important 
analytical step in understanding planetesimal dynamics.

We now provide a more detailed comparison of our results with 
previous studies and discuss the limitations of this work.

\begin{center}
\begin{deluxetable}{lcc}
\tablewidth{0pc}
\tablecaption{Different approximations for 
planetesimal dynamics in binaries}
\tablehead{
\colhead{Gravitational effects}&
\colhead{W/o gas drag}&
\colhead{With gas drag}\\
\colhead{included}&
\colhead{}&
\colhead{}
}
\startdata
Binary companion only & 2,3 & 4,5,6,8,9,10  \\
\\
Axisymmetric disk  & 7 & 1 \\
and binary companion &  &  \\
\\
Non-axisymmetric disk  & 8  & 1 \\
and binary companion &   &  \\
\enddata
\label{table}
\tablecomments{[1] This work, [2] Giuppone \etal (2011),
[3] Heppenheimer (1978), 
[4] Marzari \& Scholl (2000), [5] Paardekooper \etal (2008), 
[6] Beaug\'e \etal (2010), [7] Rafikov (2013), 
[8] Silsbee \& Rafikov (2013), [9] Thebault \etal (2006), 
[10] Xie \& Zhou (2008)}
\end{deluxetable}
\end{center}


\subsection{Comparison of different dynamical approximations.}  
\label{sect:dyn_approx}

The main novelty of our study is the extension of the line of 
analytical investigation of disk gravity effects, started in 
R13 and SR13 for axisymmetric and non-axisymmetric disks,
respectively, by including gas drag. Previous (semi-)analytical
studies of planetesimal dynamics in binaries neglected   
the gravitational effect of the disk. 

Our calculations account for both the precession of planetesimal 
orbits due to the axisymmetric part of the disk potential 
and the eccentricity excitation due to its non-axisymmetric 
component. Disk non-axisymmetry is modeled via its 
nonzero eccentricity, i.e. $m=1$ distortion, which can 
be a function of radius. We expect this approximation to 
capture the key effect of the disk asymmetry, as higher-$m$
distortions of the disk shape are relatively small 
(Marzari \etal 2012).  

In Table \ref{table} we summarize some (this list is not 
exhaustive) existing (semi-)analytical treatments of 
planetesimal dynamics (including this work), classified 
according to the physical ingredients that are taken into account.
We primarily focus on studies of secular effects to put 
our work in proper context. Our current results cover {\it all 
dynamical regimes} listed in this table in appropriate limits.  
The majority of previous studies considered
planetesimal dynamics in the presence of gas drag, with
{\it only the direct binary} gravitational perturbations taken into 
account (Marzari \& Scholl 2000; Th\'ebault \etal 2004, 
2006, 2008, 2009; Paardekooper \etal 2008). As shown 
in SR13 this approximation is unwarranted as long as 
the disk mass $M_d\gtrsim 10^{-2}M_\odot$ since then 
the disk potential dominates gravitational perturbation.  

\begin{figure}[h]
\epsscale{1.2}
\plotone{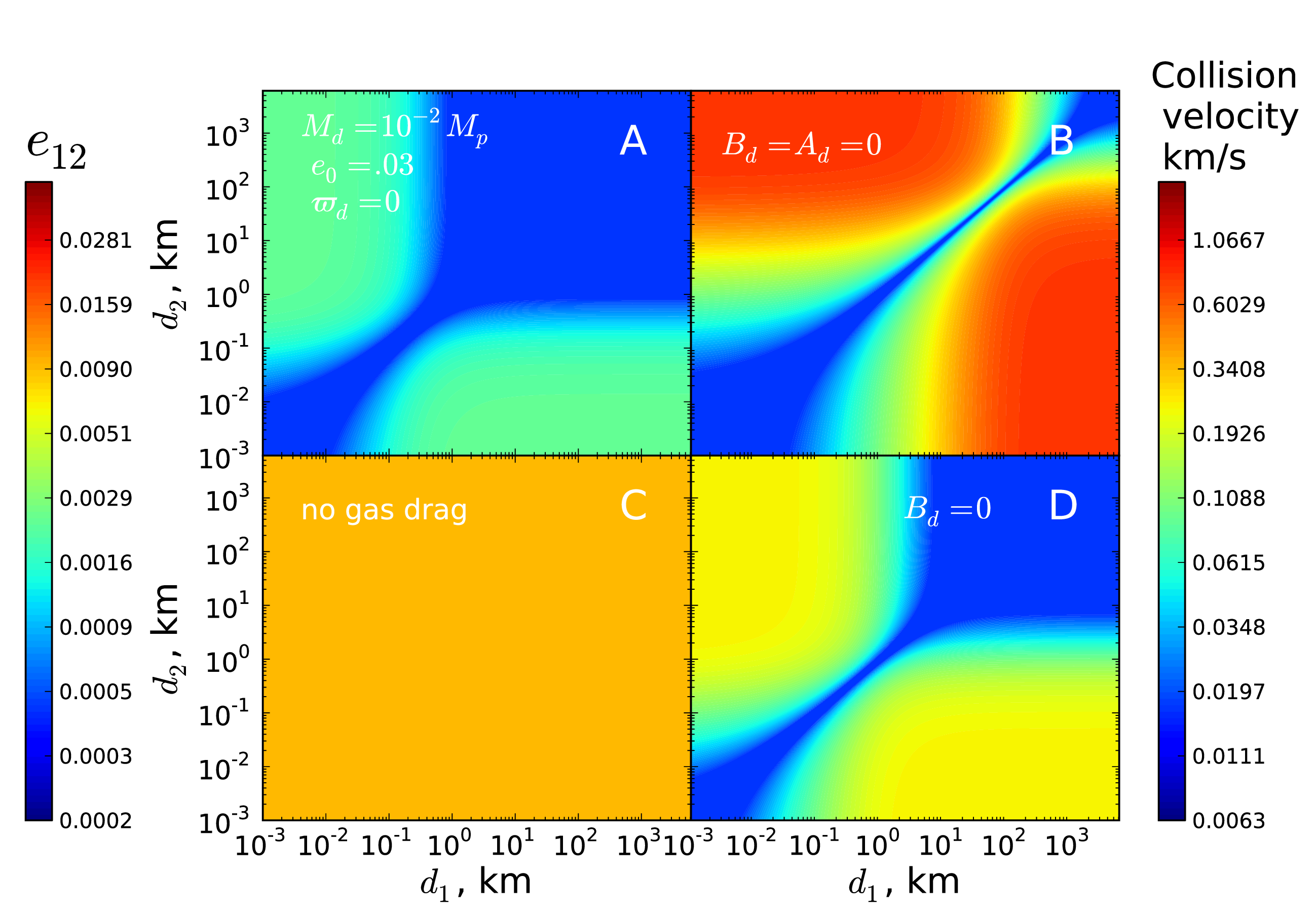}
\caption{
Comparison of different approximations for describing planetesimal 
dynamics (indicated on panels), as reflected in the map of the 
relative eccentricity of planetesimals $e_r$ of different sizes;
see text for details. Maps are drawn for an aligned disk in 
$\gamma$ Cep at 1AU (note the different semi-major axis compared 
to other figures). 
\label{fig:dif_dyn_approx}}
\end{figure}

We also provide full analytical solutions for test particle  
dynamics in a general precessing or non-precessing disk without 
companion perturbation, see equations 
(\ref{eq:e_r_pr})-(\ref{eq:d_1_pr}). Previously, 
Beaug\'e \etal (2010) studied this regime for a precessing disk 
but did not account for the gravitational effect of the disk 
(i.e. only gas drag was taken into account). 
In Figure \ref{fig:dif_dyn_approx} we illustrate the 
differences in various descriptions of planetesimal dynamics. 
It shows relative eccentricity as a function of planetesimal sizes 
$d_1$ and $d_2$ at 1 AU in an aligned disk of $M_d=10^{-2}M_p$
and $e_0=0.1$ around a primary of $\gamma$ Cephei in four 
different limits. Panel (a) presents a full calculation 
with all physical ingredients (gas drag, gravity of both the 
eccentric disk and the binary companion) accounted for using 
the solutions obtained in \S \ref{sect:ecc_evol}. 

In panel (b) we show how things change if disk gravity 
is completely switched off by setting $A_d=B_d=0$ --- an 
approximation common to a number of previous studies 
(Marzari \& Scholl 2000; Th\'ebault \etal 2004, 2006, 2008, 
2009; Paardekooper \etal 2008; Beaug\'e 
\etal 2010). One can see that without disk gravity relative 
planetesimal velocities go up by a factor of several. Moreover, the ``waist''
between the two high-$e_{12}$ regions in panel (b) is narrowest 
at $d_1\sim d_2\sim 10^2$ km, which is considerably larger than in
panel (a) where this happens for $d\sim 0.3$ km objects. 
This difference is in complete agreement with equations 
(\ref{eq:d_c_b}) and (\ref{eq:d_c_d}). 

In panel (c) we account for the gravitational effect of a 
non-axisymmetric disk but neglect gas drag ($\tau_d\to \infty$), 
i.e. use equations
(\ref{eq:e_n_drag_decompose})-(\ref{eq:e_forced_d}), as was 
done in SR13. In the absence of gas 
drag there is no apsidal alignment of planetesimal orbits 
and they approach each other at random phases. Also, $e_r$ is 
independent of $d_1$ and $d_2$ (the size-dependent drag is absent) 
explaining uniform color in Figure \ref{fig:dif_dyn_approx}c. 
Absence of gas drag results in rather high relative velocities
of planetesimals making their survival in collisions problematic.
Thus, apsidal alignment of planetesimal orbits and eccentricity 
suppression due to gas drag are very important for the proper
description of their dynamics.

Finally, in panel (d) we retain only the axisymmetric component 
of the disk potential neglecting the eccentricity excitation 
by the disk, i.e. $B_d=0$ but $A_d\neq 0$.  In this limit, also 
neglecting gas drag (accounted for here) R13 predicted dramatic 
lowering of ${\bf e}_p$. Comparison with panel (a) clearly shows 
this not to be the case when gas drag included, which can be 
understood by noticing that $e_c$ in equation (\ref{eq:e_c}) 
can significantly deviate from $\left|{\bf e}_p^{\rm n/drag}\right|$
because of ${\bf e}_g$ contribution. This is why lowering 
$\left|{\bf e}_p^{\rm n/drag}\right|$ by setting $B_d=0$ and 
increasing $|A|$ does not necessarily result in smaller $e_c$,
as expected in R13.

To summarize, simultaneously accounting for all the physical
processes affecting planetesimals --- gas drag, disk and secondary
gravity --- is very important for understanding planetesimal
growth. Omission of even a single physical ingredient can 
significantly affect the conclusions drawn from the dynamical 
calculations.

Previously Kley \& Nelson (2007) and Fragner \etal (2011) 
numerically explored planetesimal dynamics in gaseous disks, 
which were evolved using direct hydrodynamical simulations. 
They accounted for the effect of disk gravity on planetesimal 
motion and at least some of their calculations assumed 
coplanarity of the disk and the binary. However, even though 
the setup of these studies is very similar to that of our 
present work, some subtle differences prevent direct 
comparison of their results. In particular, when estimating 
the relative velocities of planetesimals based on their orbit 
crossing Kley \& Nelson (2007) do not take into account the 
apsidal phasing of their orbits (Marzari \& Scholl 2000), 
clearly obvious in their Fig. 10. As a result they find 
very high relative speeds even between equal-size 
planetesimals, which we believe is an artefact of their 
neglect of apsidal phasing. Fragner \etal (2011) study
the case of a {\it circular} binary, in which apsidal 
phasing is naturally absent, resulting in high 
relative speeds of planetesimals. As a result, 
the applicability of calculations using circular binaries
to understanding planetesimal dynamics in eccentric systems
like $\gamma$ Cep is not obvious.


\subsection{Limitations of this work.}  
\label{sect:limits}

Finally, we discuss limitations of our study. Some of them 
have to do with the adoption of secular, i.e. orbit-averaged,
approximation. While averaging over the planetesimal orbit is 
justified because $n_p^{-1}$ is always much shorter than other 
periodicities (e.g. of planetesimal apsidal precession), when 
averaging over the longer binary period one may overlook important
dynamical features of the systems possessing very massive disks. 
Indeed, equation (\ref{eq:A_d}) suggests that for 
$M_d\sim 0.1 M_\odot$ planetesimal precession rate $|A_d|$
becomes comparable to the binary angular frequency ---
$n_p\approx 0.1$ yr$^{-1}$ for $\gamma$ Cephei. In these conditions 
averaging over the latter is not justified and new effects, 
such as the possibility of evection resonance (Touma \& Wisdom 1998) 
inside the disk, may additionally affect planetesimal dynamics.  

Other effects omitted in our study, such as the density waves or 
higher-$m$ contributions to the azimuthal mass distribution 
in the disk, short-term fluctuations of the disk potential, may 
also affect planetesimal dynamics. They may account for some of 
the difference between the results of this work, which uses secular, 
time-averaged description of the disk and binary potential, 
and direct numerical studies of Kley \& Nelson (2007) and 
Fragner \etal (2011). 
Planetesimal eccentricity can be additionally excited 
by the stochastic gravitational perturbations due to the 
turbulence in the disk. This issue has been previously 
investigated for disks around single stars (Ida \etal 2008; 
Yang \etal 2009, 2012) and for circumbinary disks (Meschiari 2012).

Coplanarity of the disk and the binary orbit is another restriction,
that can be easily eliminated in future studies. We believe that 
small but 
non-zero inclination (Xie \& Zhou 2009) would not affect our solutions
for the behavior of planetesimal eccentricity. However, as shown in 
Xie \etal (2010), such non-zero inclination has strong effect 
on planetesimal collision rates.

There is also room for improvement within the framework of 
our model. Some approximations that we adopt such as the power 
law behavior of $\Sigma(a)$ and $e_d(a)$, 
constant\footnote{Variable $\varpi_d(a)$ can be used to describe disks 
with density waves.} $\varpi_d(a)$ are 
dictated by our desire to obtain analytical solutions using 
the results of SR13 whenever possible. Also, we did not 
investigate the conditions under which our model (\ref{eq:Sigma0}) 
represents a steady-state solution for a fluid disk perturbed 
by a companion (Statler 2001). More refined semi-analytical or
numerical calculations using improved disk models are certainly 
desirable but are unlikely to seriously affect our results and 
conclusions.


\section{Summary.}  
\label{sect:summ}


We studied secular dynamics of planetesimals and explored
prospects for planet formation around one of the components 
of an eccentric binary. We believe that our study includes 
most, if not all, of the important physical ingredients
relevant for this problem --- perturbations due to the binary, 
gas drag, and gravitational effects of an eccentric disk. 
This is the first time planetesimal dynamics in binaries have been 
studied analytically in such generality. The analytical nature of our 
solutions for planetesimal dynamical variables allowed 
us to explore their dependence on system parameters in 
great detail. 

Our main results can be summarized as follows:
\begin{itemize}
\item 
We find that under the action of 
gas drag as well as the gravitational effects of the binary 
companion {\it and the eccentric disk}, planetesimal eccentricity 
vector ${\bf e}_p$ converges to a constant value depending on 
the planetesimal size and the disk 
and binary properties. We obtained complete analytical solutions for
${\bf e}_p$ in the case of non-precessing disk and analyzed them 
in detail, extending results of previous studies.

\item 
We showed that relative particle-gas (equation (\ref{eq:e_r_d})) and 
particle-particle velocities can be expressed as simple functions 
of only two key parameters --- the characteristic eccentricity 
$e_c$ and planetesimal size $d/d_c$ in units of characteristic 
size $d_c$, given by equations (\ref{eq:e_c}) and (\ref{eq:d_1}).
Behavior of these variables has been explored in detail in
\S \ref{sect:pl_dyn}.

\item 
We show that in massive disks containing enough gas to form 
giant planets ($M_d\gtrsim 10^{-2}M_\odot$) planetesimal dynamics 
is always in the regime when 
{\it apsidal precession of planetesimal orbits is dominated 
by disk gravity}, i.e. in the DB or DD regimes in 
classification of SR13. Significantly eccentric ($e_0\gtrsim 
10^{-3}$) disks also dominate eccentricity excitation of 
planetesimals by their gravity (DD regime). This emphasizes  
the key role of the disk gravity in relation to planet 
formation in binaries. 

\item 
We derive the explicit form of the relative velocity distribution 
between the populations of planetesimals with different sizes 
and show that it depends only on the relative eccentricity 
$e_{12}$ of the approaching objects. 

\item 
In disks aligned with the binary planetesimals collide with 
lower velocities than in mis-aligned disks. Thus, planetesimal 
growth favors disk-binary apsidal alignment.

\item 
We also present analytical results for the dynamics of 
planetesimals in precessing disks in certain limits.

\end{itemize}

Our results will be used in Paper II to understand planet formation 
in small separation binaries, such as $\gamma$ Cep and 
$\alpha$ Cen. They can also be used to understand the
circumbinary planet formation.


\acknowledgements 

We are grateful to Jihad Touma for useful discussions.

\appendix


\section{Local approximation.}  
\label{sect:local}


Here we review local (or guiding center) approximation, 
which is often used in studies of planetesimal and 
galactic dynamics (Binney \& Tremaine 2008) 
and forms the basis of the so-called Hill approximation 
(H\'enon \& Petit 1986; Hasegawa \& Nakazawa 1990). 
In this approach eccentric motion of a 
planetesimal is considered in a locally Cartesian 
frame $(x_p,y_p)$, with ${\bf e}_x$, ${\bf e}_y$ pointing
in the radial and horizontal directions, correspondingly.
The origin of this frame is in circular Keplerian motion 
at some characteristic semi-major axis $a_0$, which is  
close to planetesimal semi-major axis $a_p$, so that 
$b_p\equiv |a_p-a_0|\ll a_p$. Equations of motion for
a particle of mass $m_p$ subject to external force 
${\bf F}=(F_x,F_y)$ can be reduced to
\ba
\ddot x_p-2n_p\dot y_p-3 n_p^2 x_p=F_x/m_p,
~~~~~~~~
\ddot y_p+2n_p\dot x_p =F_y/m_p.
\label{eq:xy_eq}
\ea

Provided that $e_p\ll 1$ one can 
represent planetesimal motion unperturbed by external 
forces as
\ba
x_p=b_p-a_0(k_p\cos n_p t+h_p\sin n_p t),
~~~~~~~~
y_p=\psi_p-\frac{3}{2}n_p b_p t +
2a_0(k_p\sin n_p t-h_p\cos n_p t),
\label{eq:unpertxy}
\ea
where $\psi_p$ is a constant and ${\bf e}_p=(k_p,h_p)$. 
This is an exact solution 
of equations (\ref{eq:xy_eq}) with 
${\bf F}=0$ and is a superposition of linear
shear and epicyclic motion.

Assuming that fluid in a gaseous disk also moves on eccentric 
Keplerian orbits, motion of the gas can be represented by
analogous equations
\ba
x_g=b_g-a_0(k_g\cos n_p t+h_g\sin n_p t),
~~~~~~~
y_g=\psi_g-\frac{3}{2}n_p b_g t +
2a_0(k_g\sin n_p t-h_g\cos n_p t).
\label{eq:g_unpertxy}
\ea

{\it Relative} motion of a particular fluid element and a
particle is described using relative coordinates 
$x_r=x_p-x_g$, $y_r=y_p-y_g$. According to equations 
(\ref{eq:unpertxy})
\ba
x_r=b_r-a_p(k_r\cos n_p t+h_r\sin n_p t),
~~~~~~~
y_r=\psi_r-\frac{3}{2}n_p b_r t +
2a_p(k_r\sin n_p t-h_r\cos n_p t),
\label{eq:rel_unpertxy}
\ea
where $k_r\equiv k_p-k_g$, $h_r\equiv h_p-h_g$ are the components
of the relative eccentricity vector, $b_r\equiv b_p-b_g$ is
the semi-major axis separation between the particle and fluid 
element, and $\psi_r\equiv \psi_p-\psi_g$. We have also 
used the fact that $a_g\approx a_0\approx a_p$ and switched
from $a_0$ to $a_p$. 

Velocity of Keplerian motion in the local approximation is
obtained by differentiating equations  
(\ref{eq:rel_unpertxy}) with respect 
to time. In particular, relative particle-gas velocity is 
given by 
\ba
v_{x,r}=n_p a_p(k_r\sin n_p t-h_r\cos n_p t),
~~~~~~~~
v_{y,r}=-\frac{3}{2}n_p b_r +
2n_p a_p(k_r\cos n_p t+h_r\sin n_p t).
\label{eq:v_unpertxy}
\ea

Analogous formulae apply to the relative motion of two planetesimals 
with sizes $d_1$ and $d_2$, with the replacement $e_r\to e_{12}$, 
$b_r\to b_{12}$, $(x_r,y_r)\to (x_{12},y_{12})$, and so on. 
In particular, equation (\ref{eq:rel_unpertxy}) shows that 
two objects with $|b_{12}|<a_p e_{12}$ can experience close 
approaches. When this happens $x_{12}=y_{12}=0$ and $b_{12}$ 
can be eliminated from equation (\ref{eq:v_unpertxy}) giving
\ba
v_{12,y}(x_{12}=0)=\frac{1}{2}n_p a_p(k_{12}\cos n_p t+h_{12}\sin n_p t),
\label{eq:v_unperty1}
\ea
(here ${\bf e}_{12}=(k_{12},h_{12})$) so that the relative 
approach velocity (i.e. the velocity unaffected by the mutual 
gravitational attraction of particles) is 
\ba
v_{12}=n_p a_p\left[k_{12}^2+h_{12}^2-(3/4)(k_{12}\cos n_p t
+h_{12}\sin n_p t)^2\right]^{1/2}.
\label{eq:v_unper}
\ea

Whenever particle is affected by forces other than the stellar
gravity, i.e. ${\bf F}\neq 0$, solutions (\ref{eq:unpertxy}) 
are no longer strictly valid. However, one can still represent
particle motion via these solutions, assuming that 
orbital elements {\it osculate}, i.e. evolve in time.
Hasegawa \& Nakazawa (1991) derived equations for the 
orbital element evolution, in particular 
\ba
\dot a_p=\dot b_p=\frac{2F_y}{n_p m_p},
~~~~\dot k_p=\frac{1}{n_p a_p m_p}\left(2F_y\cos n_p t+
F_x\sin n_p t\right),
~~~~
\dot h_p=\frac{1}{n_p a_p m_p}\left(2F_y\sin n_p t-
F_x\cos n_p t\right).
\label{eq:kh_dot}
\ea
For a given force expression ${\bf F}$ these equations, after 
averaging over the orbital period, represent the extra terms 
entering the equations (\ref{eq:dhdt_gen})-(\ref{eq:dkdt_gen}).


\section{Planetesimal eccentricity in a precessing disk in the 
case of linear drag.}  
\label{sect:lin_t_sol}


Here we derive the full time-dependent solution for 
planetesimal eccentricity starting with arbitrary initial 
conditions and assuming that the gas drag is {\it linear}, 
i.e. $\tau_d$ in equations (\ref{eq:hdrag1}) is a constant 
independent of ${\bf e}_p$. We also include a possibility of 
the uniform disk precession so that 
$\varpi_d(t)=\dot\varpi_d t+\varpi_{d0}$. Then equations 
(\ref{eq:dhdt_gen})-(\ref{eq:dkdt_gen}) represent
a linear system of equations which can be trivially solved
to give
\ba
\left\{
\begin{array}{l}
k(t)\\
h(t)
\end{array}
\right\}
=e_{\rm free}e^{-t/\tau_d}
\left\{
\begin{array}{l}
\cos\left(At+\varpi_0\right)\\
\sin\left(At+\varpi_0\right)
\end{array}
\right\}
+
\left\{
\begin{array}{l}
k_{\rm f}\\
h_{\rm f}
\end{array}
\right\},
\label{eq:t_dep}
\ea
where the first term represents the free eccentricity, with 
$e_{\rm free}$ and $\varpi_0$ being constant, while the second is
the forced eccentricity 
${\bf e}_f = (k_{\rm f},h_{\rm f})= {\bf e}_{f,b}+{\bf e}_{f,d}$, where 
${\bf e}_{f,b}$ is given by equation (\ref{eq:e_f_b}) and 
\ba
{\bf e}_{f,d} =  
\left[\frac{e_g^2+\tau_d^2 B_d^2}{1+\tau_d^2
\left(A-\dot\varpi_d\right)^2}\right]^{1/2}
\left\{
\begin{array}{l}
\cos\left(\varpi_d(t)+\phi\right)\\
\sin\left(\varpi_d(t)+\phi\right)
\end{array}
\right\},~~~~~
\cos\phi=\frac{e_g-\tau_d^2 B_d\left(A-\dot\varpi_d\right)}
{\left(e_g^2+\tau_d^2 B_d^2\right)^{1/2}
\left[1+\tau_d^2\left(A-\dot\varpi_d\right)^2
\right]^{1/2}}.
\label{eq:phi}
\ea
In the limit of slow precession $|\dot \varpi_d|\ll |A|$ one
finds that ${\bf e}_f$ is given by expressions 
(\ref{eq:vect_sum})-(\ref{eq:phi_nonprec}). Generally, the relative 
planetesimal-gas eccentricity ${\bf e}_r={\bf e}_f-{\bf e}_g$ is
\ba
{\bf e}_r = {\bf e}_{f,b} -
\tau_d\frac{B_d+e_g\left(A-\dot\varpi_d\right)}
{\left[1+\tau_d^2\left(A-\dot\varpi_d\right)^2\right]^{1/2}}
\left\{
\begin{array}{l}
\cos\left(\varpi_d(t)-\phi_r\right)\\
\sin\left(\varpi_d(t)-\phi_r\right)
\end{array}
\right\},~~~~~
\cos\phi_r=\frac{\tau_d\left(A-\dot\varpi_d\right)}
{\left[1+\tau_d^2\left(A-\dot\varpi_d\right)^2
\right]^{1/2}}.
\label{eq:ph_r}
\ea

The first forced term ${\bf e}_{f,b}$ results from excitation 
by the binary companion. It is constant in time and is independent 
of $\dot\varpi_d$. 
The second term is induced by the disk via both its gravitational 
potential and gas drag. This contribution to ${\bf e}_f$
circulates at the disk precession frequency $\dot \varpi_d$ and 
its amplitude is sensitive to $\dot\varpi_d$. 

Independent of the initial conditions (i.e.
the values of $e_{\rm free}$ and $\varpi_0$) the free eccentricity
contribution damps out on a characteristic timescale $\tau_d$.
As a result, in the long run ${\bf e}_p$ inevitably converges 
to ${\bf e}_f$. 

In the limit of strong gas drag, $\tau_d\to 0$, one finds that 
${\bf e}_f\to {\bf e}_g$ as expected, since drag is strong 
enough to align planetesimal orbits with fluid trajectories. 
In this limit the relative eccentricity between planetesimals 
of different sizes having different damping times $\tau_{d,1}$ 
and $\tau_{d,2}$ is
\ba
e_{12}\to \left|A-\dot\varpi_d\right|
\left|\tau_{d,1}-\tau_{d,2}\right|
\left[\left(e_c^{\rm pr}\sin\varpi_d\right)^2+
\left(e_c^{\rm pr}\cos\varpi_d
+k_b\frac{A}{A-\dot\varpi_d}\right)^2\right]^{1/2},~~~
\tau_{d,1},\tau_{d,2}\ll |A-\dot\varpi_d|^{-1},
\label{eq:e_r_strong}
\ea
where $e_c^{\rm pr}$ is given by equation (\ref{eq:e_prec}).

In the opposite extreme $\tau_d\to \infty$ (weak drag) one 
finds $\phi\to \pi$ and ${\bf e}_f$ reduces to the forced 
eccentricity value (with disk precession) obtained in 
SR13. The relative velocity becomes
\ba
e_{12}\to \frac{1}{\left|A-\dot\varpi_d\right|}
\left|\tau_{d,1}^{-1}-\tau_{d,2}^{-1}\right|
\left[\left(e_c^{\rm pr}\sin\varpi_d\right)^2+
\left(e_c^{\rm pr}\cos\varpi_d
+k_b\frac{A-\dot\varpi_d}{A}\right)^2\right]^{1/2},~~~
\tau_{d,1},\tau_{d,2}\gg |A-\dot\varpi_d|^{-1}.
\label{eq:e_r_weak}
\ea
Note that in this expression $k_b$ is multiplied by a factor 
different from that in equation (\ref{eq:e_r_strong}). However,
it is clear that in both limiting cases $e_{12}\ll e_1,e_2$,
i.e. the {\it relative} planetesimal eccentricity is much 
less than the individual eccentricities $e_1$ and $e_2$, a 
result that remains valid in a precessing disk.


\section{Planetesimal eccentricity in a precessing disk in the 
case of quadratic drag.}  
\label{sect:quad_t_sol}

\begin{figure}
\vspace{-1.5cm}
\plotone{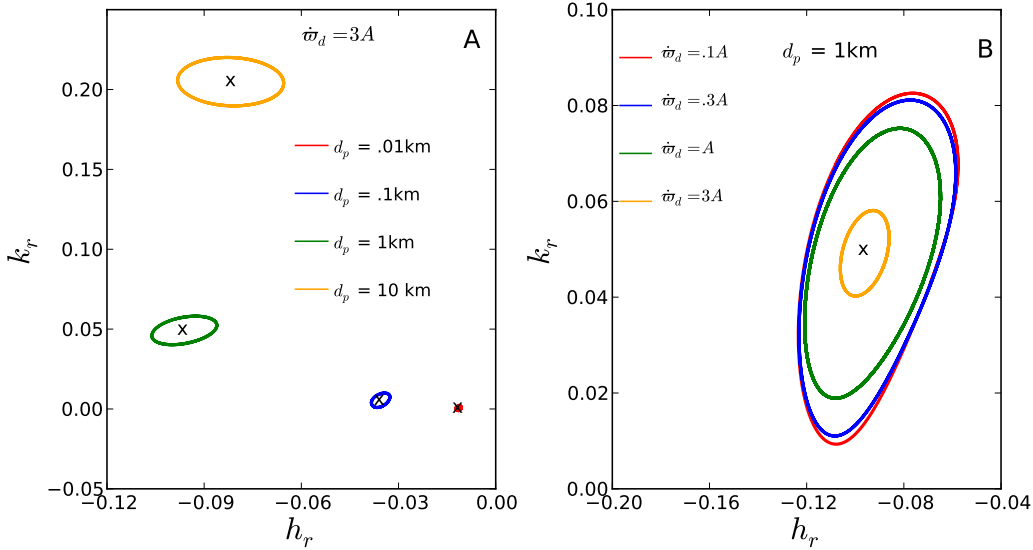}
\caption{
Limit cycles to which relative gas-planetesimal eccentricity 
vector ${\bf e}_r$ 
converges in precessing disks. Panel (a) shows evolution of the 
limits cycles as a function of planetesimal size $d_p$, while in
panel (b) we vary disk precession rate $\dot\varpi_d$. Calculations
have been performed at $a_p=2.5$ AU in a standard aligned disk with 
$M_d = 10^{-3} M_p$, $e_0 =0.04$, $a_{\rm out}= 5$ AU in
$\gamma$ Cep system. These parameters place planetesimal dynamics 
in the strong binary perturbation regime, see \S 
\ref{sect:precess_bin}. Crosses mark the centers of the limit
cycles computed according to equation (\ref{eq:strong_bin}). 
Note the evolution of the positions and shapes of the limit cycles 
as $d_p$ and $\dot\varpi_d$ are varied.
\label{fig:limit_cycles}}
\end{figure}

In the case of quadratic drag (\ref{eq:F_drag}) Figure
\ref{fig:ev_traj}b clearly shows the phenomenon of ${\bf e}_p$ 
convergence to a quasi-stationary limit cycle behavior, 
similar to the results of \S \ref{sect:lin_t_sol}.
This behavior is further illustrated in Figure 
\ref{fig:limit_cycles}, where we show the dependence 
of the limit cycles on planetesimal size $d_p$ and disk
precession rate $\dot\varpi_d$. Because 
of the nonlinear drag law the shapes of the limit cycles 
in general deviate from ellipses. 

Nevertheless, their
gross features still can be understood our linear solution
(\ref{eq:ph_r}). In particular, limit cycles are not 
centered on $(k_r,h_r)=0$ because of the binary companion 
perturbations, i.e. non-zero ${\bf e}_{f,b}$ varying as 
$d_p$ (and $\tau_d$) change. Amplitude of the limit cycles
goes down for smaller $d_p$ because $\tau_d$ is also smaller,
which according to equation (\ref{eq:ph_r}) reduces the
oscillating contribution to ${\bf e}_r$. As we vary 
$\dot\varpi_d$ in Figure \ref{fig:limit_cycles}b the binary 
contribution stays unchanged and all limits cycles stay 
centered on the same point in $h_r$-$k_r$ space. 

Their sizes ary with $\dot\varpi_d$ as predicted by 
equation (\ref{eq:ph_r}). They shrink at high 
$|\dot\varpi_d|\sim |A|$, in agreement with equation (\ref{eq:ph_r}).
For slow precession $|\dot\varpi_d|\ll |A|$ limit cycles 
converge to the trajectory for the non-precessing disk 
solution (\ref{eq:vect_sum}) in which $\varpi_d$ is set to
vary as $\varpi_d(t)=\dot\varpi_d t+\varpi_{d0}$. Note that 
such convergence to solution (\ref{eq:vect_sum}) is obvious 
only in the case of $|\dot\varpi_d \tau_d|\ll 1$, i.e. when 
gas drag allows ${\bf e}_p$ to quickly readjust to a new 
``quasi-static'' solution as $\varpi_d$ changes. This is the 
case shown in Figure \ref{fig:limit_cycles}b. In the
opposite case of $|\dot\varpi_d \tau_d|\gg 1$ (and
$|\dot\varpi_d|\ll |A|$) this convergence is not obvious 
as the disk precession constantly drives free eccentricity, 
while the gas drag is not strong enough to quickly damp it. 
We leave detailed exploration of such details to a future 
study.

Now, let us rewrite equations 
(\ref{eq:dhdt_gen})-(\ref{eq:dkdt_gen}) in terms of the 
relative particle-gas eccentricity components $k_r$ and $h_r$:
\ba
&&\frac{dh_r}{dt}=Ak_r-\frac{h_r}{\tau_d}+B_b+
\left[\left(A-\dot\varpi_d\right)e_g+B_d\right]\cos\varpi_d(t),
\label{eq:dhdt_rel}\\
&&\frac{dk_r}{dt}=-Ah_r-\frac{k_r}{\tau_d}-
\left[\left(A-\dot\varpi_d\right)e_g+B_d\right]\sin\varpi_d(t),
\label{eq:dkdt_rel}
\ea
with $\tau_d$ given by equation (\ref{eq:tau_d}) and 
dependent upon $e_r$.

Explicit time dependence of the last terms in these nonlinear 
equations precludes us from finding their 
general analytical solutions even in the case of the 
limit-cycle behavior. However, we can still obtain analytical 
results for planetesimal eccentricity in the two limiting cases, 
reviewed next.

First, one can assume that binary companion dominates eccentricity 
forcing, which implies that the condition (\ref{eq:bin_dom_cond}) 
is fulfilled. Then one can drop last $\varpi_d$-dependent terms 
in equations (\ref{eq:dhdt_rel})-(\ref{eq:dkdt_rel}) removing 
the explicit time dependence from them. This is essentially 
equivalent to neglecting both the gravitational effect of the disk,
i.e. $|{\bf e}_d|\to 0$, and the gas eccentricity $e_g$ compared to 
$|{\bf e}_b|$. As a result, we find a steady-state solution 
(\ref{eq:strong_bin}) for $k_r\approx k_p$ and $h_r\approx h_p$, 
which is essentially the equations (\ref{eq:k_p})-(\ref{eq:h_p}) with 
$|{\bf e}_d|,|{\bf e}_g|\to 0$. Then planetesimal dynamics 
is described by the analytical results of \S \ref{sect:ecc_evol} 
with $e_c\approx |B_b/A|$.

In the opposite extreme of weak eccentricity excitation by the 
binary companion we introduce new coordinates 
$H\equiv k_gh_r-h_gk_r$, $K\equiv h_gh_r+k_gk_r$ (see 
Beaug\'e \etal 2010 for a similar treatment). Then the 
evolution of $H$ and $K$ is given by 
\ba
&&\frac{dH}{dt}=\left(A-\dot\varpi_d\right)K-\frac{H}{\tau_d}+B_bk_g(t)+
e_g\left[\left(A-\dot\varpi_d\right)e_g+B_d\right],
\label{eq:dHdt_rel}\\
&&\frac{dK}{dt}=-\left(A-\dot\varpi_d\right)H-\frac{K}{\tau_d}+B_bh_g(t).
\label{eq:dKdt_rel}
\ea
When the eccentricity excitation by the companion is small 
we can drop the $B_b$ terms in these equations, removing the 
explicit time-dependence, which appears because of circulating 
$k_g$ and $h_g$. As a result, we find the steady state 
solutions for $H$ and $K$ in the implicit form
\ba
K=e_c^{\rm pr}e_g\frac{\left(A-\dot\varpi_d\right)^2\tau_d^2}
{1+\left(A-\dot\varpi_d\right)^2\tau_d^2},~~~~~
H=-e_c^{\rm pr}e_g\frac{\left(A-\dot\varpi_d\right)\tau_d}
{1+\left(A-\dot\varpi_d\right)^2\tau_d^2},
\label{eq:steady}
\ea
where $\tau_d$ is a function of the relative particle-gas 
eccentricity $e_r=e_g^{-1}\left(K^2+H^2\right)^{1/2}$ given by 
equation (\ref{eq:e_r_pr}). This solution corresponds to 
eccentricity vector ${\bf e}_p$ fixed in a disk frame, which 
uniformly precesses at the rate $\dot \varpi_d$. 

Using these solutions it
is trivial to show that ${\bf e}_p\to {\bf e}_{f,d}$ given by
equation (\ref{eq:phi}) with $B_b$ set to zero. That in the 
weak binary perturbation regime we find the same expression 
for ${\bf e}_p$ as in the case of linear drag is not surprising:  
with $B_b=0$ one finds that $|{\bf e}_r|$ is constant in time, 
so that $\tau_d$ is also constant. Then equations 
(\ref{eq:dHdt_rel})-(\ref{eq:dKdt_rel}) are the same as in 
the linear drag case and have the same 
steady state solutions (\ref{eq:steady}).

\end{document}